\documentclass[11pt]{article}
\usepackage{multirow}
\usepackage{amssymb}
\usepackage{amsmath}
\usepackage{amsthm}
\usepackage{amsfonts}
\usepackage{xcolor}
\usepackage{mathbbol}
\usepackage{bm}
\usepackage{hyperref}
\usepackage{cleveref}
\usepackage{ dsfont }

\usepackage{times}
\usepackage{graphicx}
\usepackage{bm}

\usepackage{dsfont}
\RequirePackage[algo2e]{algorithm2e}
\RequirePackage{enumitem}
\RequirePackage{algorithm}
\usepackage{titling}

\usepackage[numbers]{natbib}

\usepackage[utf8]{inputenc}
\usepackage{setspace}
\usepackage[margin=1in]{geometry}
\usepackage{authblk}
\title{Adaptive Projected Two-Sample Comparisons \\for Single-Cell Gene Expression Data}
\author{Tianyu Zhang}
\author{Jing Lei}
\author{Kathryn Roeder}
\affil{Department of Statistics and Data Science, Carnegie Mellon University}
\date{\vspace{-5ex}}

\usepackage[page]{appendix}
\newtheorem{theorem}{Theorem}[section]
\newtheorem{lemma}[theorem]{Lemma}
\newtheorem{corollary}[theorem]{Corollary}

\newtheorem{assumption}[theorem]{Assumption}
\newtheorem{remark}{Remark}[section]

\begin{document}

\maketitle
\begin{abstract}
We study high-dimensional two-sample mean comparison and address the curse of dimensionality through data-adaptive projections. Leveraging the low-dimensional and localized signal structures commonly seen in single-cell genomics data, our first proposed method identifies a sparse, informative low-dimensional subspace and then performs statistical inference restricted to this subspace. To address the double-dipping issue---arising from using the same data for projection and inference---we develop a debiased projected estimator using the semiparametric double-machine learning framework. The resulting inference not only has the usual frequentist validity but also provides useful information on the potential location of the signal due to the sparsity of the projection.
Our second method uses a more flexible projection scheme to improve the power against the global null hypothesis and avoid the degeneracy issue commonly faced by existing methods. It is particularly useful when debiasing is practically challenging or when the informative signal is not well-captured by the subspace. Experiments on synthetic data and real datasets demonstrate the theoretical promise and interpretability of the proposed methods.  
\end{abstract}


\section{Introduction}\label{section: intro}

Comparing the mean vectors of two high-dimensional random vectors is a canonical statistical problem with applications in science, engineering, and business. The problem traces back to its low-dimensional counterpart, notably Hotelling's $T^2$ \citep{hotellings} introduced in the 1930s. The high-dimensional two-sample mean comparison problem has been extensively studied in the statistical literature. See, for example, \cite{bai1996effect, chen2010two, tony2014two}. Various methods have been proposed under differing assumptions on the underlying signal structure: see \cite{huang2022overview} for a recent review and extensive numerical comparisons.

\begin{figure}[!htbp]
    \centering
    \includegraphics[width =0.8\linewidth]{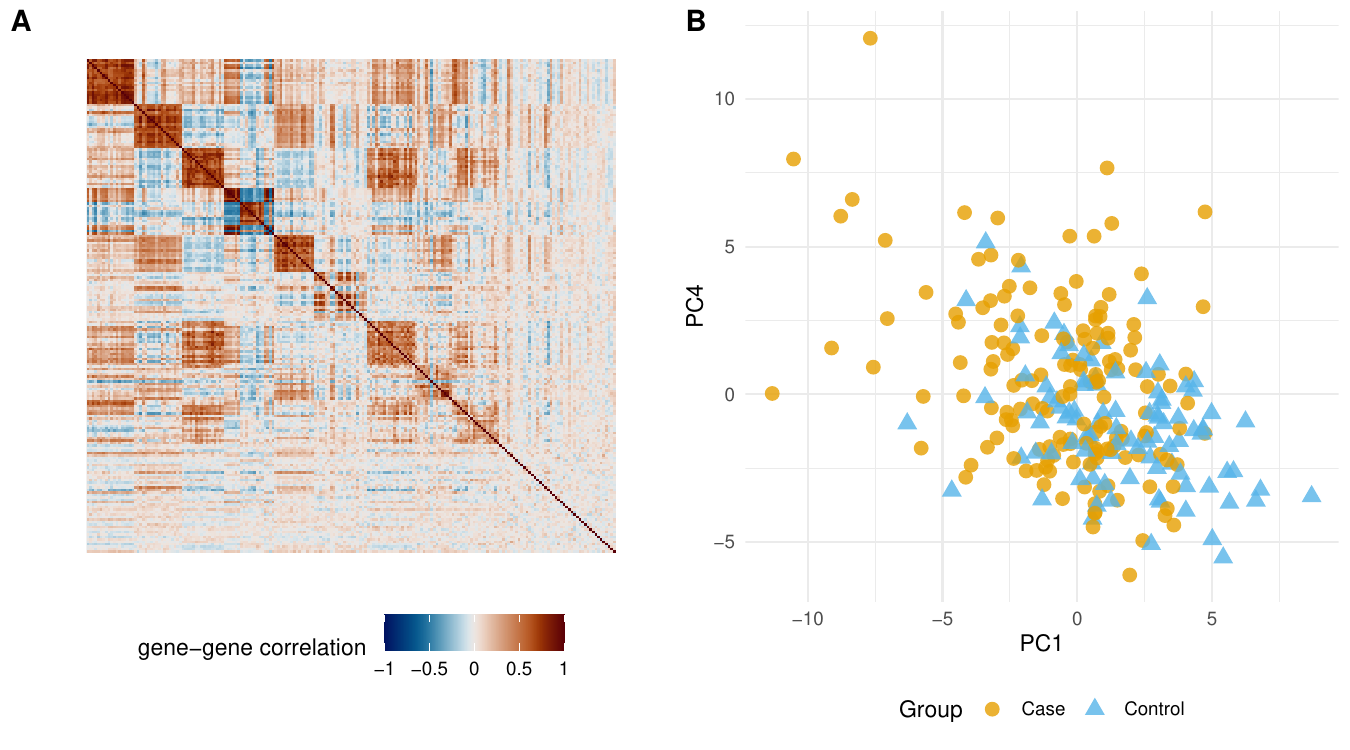}
    \caption{Co-regulation of genes and projection behavior in a T4 immune cell data.
    \textbf{A}. Sample correlations between genes in a T4 immune cell dataset. Each diagonal block corresponds to a group of genes with non-zero loadings in an estimated sparse PC. Each gene is shown only once, even if it appears in multiple PCs.
    \textbf{B}. Projection scores of each sample onto the directions specified by PC1 and PC4. PC2 and PC3 are not plotted because the distributions are visually overlapping. See Section~\ref{section: real data} for further details.}
    \label{fig: motivation}
\end{figure}

We study the problem of high-dimensional two-sample mean inference in the context of high-throughput single-cell RNA sequencing (scRNA-seq) data. Since the initial breakthrough \cite{tang2009mrna}, scRNA-seq has enabled significant advances in understanding cellular composition and gene regulatory interactions. However, the high dimensionality and the complex interaction among genes pose new challenges to existing inference methods. In particular, most existing methods output either a global $p$-value or a long list of gene-level $p$-values, providing little structural insights about the signal---such as regulatory pathways contributing to the observed difference. In practice, scientists are often interested not only in whether the two groups differ, but also in identifying the subsets of gene clusters most responsible for the difference.

In this work, we develop interpretable two-sample mean comparison procedures that provide valid inference as well as information on the signal location. Our approach is inspired by a key structural property commonly observed in scRNA-seq data: The high-dimensional gene expression difference between two groups is mostly carried by a small subset of highly correlated genes \citep{Lucas:2010, Stein-OBrien:2018}. After identifying such a small subset of genes, we can project the high-dimensional data onto a linear subspace involving only genes in this subset, reducing the high-dimensional problem to a low-dimensional one. In practice, the gene subset and subspace must be estimated from data, typically using sparse Principal Component Analysis (sPCA) or methods tailored to scRNA-seq structure \citep{langfelder2008wgcna, su2023cell}. To avoid the double-dipping issue caused by using the same data for both projection estimation and mean testing, we develop a semiparametric one-step estimator to remove the potential bias carried by the estimated projection direction. In case the mean difference is not well-aligned with the estimated projection direction, we further develop an ``anchored projection'' that enjoys both statistical validity under the null and improved power under the alternative.

As an example, we present the principal component (PC) gene clusters from a lupus dataset in (\Cref{fig: motivation}, see \Cref{section: real data} for details). In subplot A, we can observe that the number of correlation clusters is much smaller than the number of total genes \cite{stuart2003gene}. In fact, a common initial visualization with a fresh scRNA-seq dataset is a scatter plot of each sample's PC projection score, depicted in \Cref{fig: motivation}B. The projection directions are the estimated first and fourth PC directions. We observe a bimodal pattern in both the PC1 and PC4 directions, indicating that the genes contained in these two PCs may have different expression levels and are worth further investigation. It is then natural to ask whether this bimodality is due to true group differences or driven by randomness. Because the PC directions are estimated from the same data, their estimation variability must be properly accounted for. A primary goal of this work is to provide a statistically principled framework to answer this question.

Our theoretical and methodological contributions are summarized below.
\begin{itemize}
    \item We propose an adaptive projection framework for interpretable high-dimensional two-sample mean comparison.
    \item For the sPCA projection, we identify and implement the semiparametric one-step mean estimator to remove potential bias in the sparse PC projection estimate.  To our knowledge, this is the first one-step estimator using the sparse PC as a nuisance parameter. The derivation of the corresponding influence function and the proof of asymptotic negligibility of the first-order bias are both novel and technically nontrivial.
    \item Under the global null---when the two high-dimensional means are identical---debiasing is not necessary to achieve asymptotic normality. We generalize this phenomenon and provide a sufficient condition called \emph{approximate orthogonality}. This framework allows the implementation of black-box correlation discovery algorithms and/or supervised classifiers to improve power against the global null. The resulting method, called the \emph{anchored projection test}, can avoid (i) loss of power when the mean difference is not well-aligned with the sparse PC, and (ii) degeneracy of the classifier under the global null. 
\end{itemize}

\paragraph{Related work}

Explicitly relating the mean difference and correlation structure reflects the consensus that gene expressions in a cell are co-regulated. Correlated gene expression patterns often identify sparse sets of genes that control key biological systems, such as coordinated transcriptional regulation \citep{Lucas:2010, Stein-OBrien:2018}. For example, a transcription factor may regulate a set of genes sharing common features (motifs), thereby establishing a regulatory network \citep{Carvalho:2008}. Building on this idea, \cite{Knowles:2011} argued that statistically derived factors---such as PCs---frequently capture coordinated biological activity that can be usefully modeled. In a related direction, a collection of methods termed \textit{contrastive dimension reduction} \cite{zou2013contrastive, abid2018exploring, jones2022contrastive} have been developed to identify systematic differences in covariance matrices between groups of genes. 
In the mean inference literature, a recent work \cite{zhou2023new} develops a Bayesian method under a low-dimensional sparse factor model and demonstrates its ability to localize subsets of genes driving group differences. 

To establish approximately normal test statistics, we implement a semiparametric one-step procedure \citep{bickel1993efficient,tsiatis2006semiparametric, kosorok2008introduction, cher2018doubleml,kennedy2022semiparametric,hines2022demystifying,rakshit2023sihr} to construct an asymptotically normal estimate of the projection score of interest. Specifically, we leverage the influence function of PCs to reduce the bias from sPCA. This method is detailed in \Cref{section: debiased}. In \Cref{section: app orthogonality}, we discuss an alternative strategy to achieve asymptotically Gaussian test statistics, potentially incorporating supervised linear classifiers to enhance testing power. A challenge is that such a linear discriminating direction is not well-defined under the global null. This issue frequently arises in two-sample testing problems involving nuisance parameters \citep{liu2022multiple, williamson2023general, dai2022significance, lundborg2022projected}. To tackle this challenge, we develop the anchored projection test that adaptively combines the linear discriminating direction and the principal component projection.

\paragraph{Notation.} For a positive integer $M$, let $[M] = \{1,..., M\}$. We define $a \vee b = \max\{a,b\}$ and $a \wedge b = \min \{a,b\}$. Our method uses the one-step estimation framework, which involves sample-splitting and cross-fitting \citep{langfelder2008wgcna, zhou2023new, jin2020vivo}. A fraction of the data is used to estimate nuisance parameters---such as PC vectors or discriminative directions---while the remaining samples are used to construct the test statistics. Let $N_X$ and $N_Z$ denote the total sample sizes in the control and treatment groups, respectively. The integer $M$ denotes the number of folds of sample splitting. We assume the sample sizes in each fold, $n_X = N_X/M, n_Z = N_Z/M$, are integers. The data in the $m$-th fold, denoted $\mathcal{D}^{(m)}$, consists of the subsample
\begin{equation*}
\mathcal{D}^{(m)}=\left\{X_i, i=(m-1) n_X+1, \ldots, m n_X\right\}\cup\left\{Z_i, i=(m-1) n_Z+1, \ldots, m n_Z\right\}\,,
\end{equation*}
where $X_i$ and $Z_i$ denote control and treatment observations, respectively. The complete dataset is $\mathcal{D}:= \cup_{m=1}^M \mathcal{D}^{(m)}$, and the samples not in fold $m$ are denoted as $\mathcal{D}^{(-m)} = \mathcal{D} \backslash \mathcal{D}^{(m)}$. We define $n := n_X \wedge n_Z$ as the smaller per-fold sample size across groups.

For a matrix $\Sigma$, we let $\Sigma^{+}$ denote its Moore–Penrose pseudoinverse and $|\Sigma|$ its operator norm. If $\Sigma$ is positive semidefinite, we write $\lambda_i(\Sigma)$ for its $i$-th largest eigenvalue. For a random vector $\mathbb{X}$, we use 
$\Sigma_{\mathbb{X}}$ to denote the population covariance matrix of $\mathbb X$. We write $(\lambda_j,v_j)$ for its $j$-th eigenvalue and eigenvector, assuming they are uniquely defined up to a sign flip for $v_j$.

\section{Debiased Projection for the Projected Null}\label{section: debiased}

Assume we have two IID samples $\{Z_i: i \in [N_Z]\},~\{X_j: j \in [N_X]\}\subset \mathbb{R}^p$ from the case and control group, distributions $P_Z, P_X$, respectively (A more rigorous discussion regarding the high-dimensional setting will be presented in \Cref{section: debiased test}.)

We are interested in testing the following \emph{projected null hypothesis}, which is inspired by the correlation structure in scRNA data (Figure~\ref{fig: motivation}):
\begin{equation}\label{eq:proj_null}
H_0^{\rm proj}(u):\left(\mu_X-\mu_Z\right)^{\top} u=0,
\end{equation}
where $u\in\mathbb{R}^p$ is a sparse vector, and $\mu_X, \mu_Z \in \mathbb{R}^p$ are the population means of $P_X, P_Z$. In general, $u$ can be any meaningful direction determined by the model and the background knowledge. To make our discussion concrete, in this section, we focus on the case $u=v_1$, the leading PC of the shared covariance matrix $\Sigma=\Sigma_X=\Sigma_Z$.

When the vector $u$ is known, the problem \eqref{eq:proj_null} is just a simple two-sample mean test and can be effectively solved using standard methods. The vector $u$ provides both dimension reduction and, when $u$ is sparse, variable selection. In this case, rejecting the projected null $H_0^{\rm proj}(u)$ not only asserts that $\mu_X$ and $\mu_Z$ are different but also indicates that the difference has non-zero inner product with $u$. When $u$ is sparse, we can further deduce that $\mu_{X,j}\neq\mu_{Z,j}$ for some $j\in{\rm supp}(u)$, where ${\rm supp}(u)=\{j\in[p]:u_j\neq 0\}$ denotes the support of $u$.

\subsection{A Plug-in Proposal}\label{section: plug-in test}
The problem becomes more complicated when $u$ is unknown and needs to be estimated from data. 
We begin our discussion with the following plug-in test statistic that uses cross-fitting to avoid the double-dipping issue.

\begin{equation}
\label{eq:simple statistics}
    T_{\rm pi}(u)  = \hat\sigma_{\rm pi}^{-1}\sum_{m=1}^M \hat \theta_{\rm pi}^{(m)}(u)\,,
\end{equation}
where
\begin{equation*}
\hat{\theta}_{\mathrm{pi}}^{(m)}(u) = \left(\mu_X^{(m)}-\mu_Z^{(m)}\right)^{\top} u^{(-m)}\,.
\end{equation*}
Here $\mu_X^{(m)} = n_X^{-1}\sum_{X_i \in \mathcal{D}^{(m)}} X_i$, $\mu_Z^{(m)} = n_Z^{-1}\sum_{Z_i \in \mathcal{D}^{(m)}} Z_i$ are the estimated mean vectors using samples in $\mathcal{D}^{(m)}$, $u^{(-m)}$ is an estimated version of $u$ using $\mathcal{D}^{(-m)}$,
and $\hat\sigma_{\rm pi}$ is an estimate of the standard deviation of $\sum_{m=1}^M\hat\theta_{\rm pi}^{(m)}(u)$.

A variance estimator of $T_{\rm pi}(u)$ one may consider is:
\begin{equation}\label{eq: cross-fitting variance}
\hat \sigma_{\rm pi}^2 = 
\sum_{m=1}^M \left\{n_X^{-1}{\rm Var}^{(m)}(X) +  
n_Z^{-1}{\rm Var}^{(m)}(Z)\right\},
\end{equation}
where
\begin{equation}
    {\rm Var}^{(m)}(X) = n_X^{-1}\sum_{X_i \in \mathcal{D}^{(m)}}\left(X_i^{\top} u^{(-m)}-\mu_X^{(m)^{\top}} u^{(-m)}\right)^2
\end{equation}
and $ {\rm Var}^{(m)}(Z)$ is similarly defined.

The statistics $\hat{\theta}_{\mathrm{pi}}^{(m)}(u)$ is a cross-fitted projection of the mean difference. When the projected null $H_0^{\mathrm{proj}}(u)$ \eqref{eq:proj_null} is violated, we should expect the studentized statistic $T_{\mathrm{pi}}(u)$ to have a larger absolute value. However, the distribution of $T_{\mathrm{pi}}(u)$ under the null hypothesis \eqref{eq:proj_null} is not always close to the standard normal distribution due to the variability of $u^{(-m)}$.

\begin{figure}[!t]
    \centering     \includegraphics[width =0.75\linewidth]{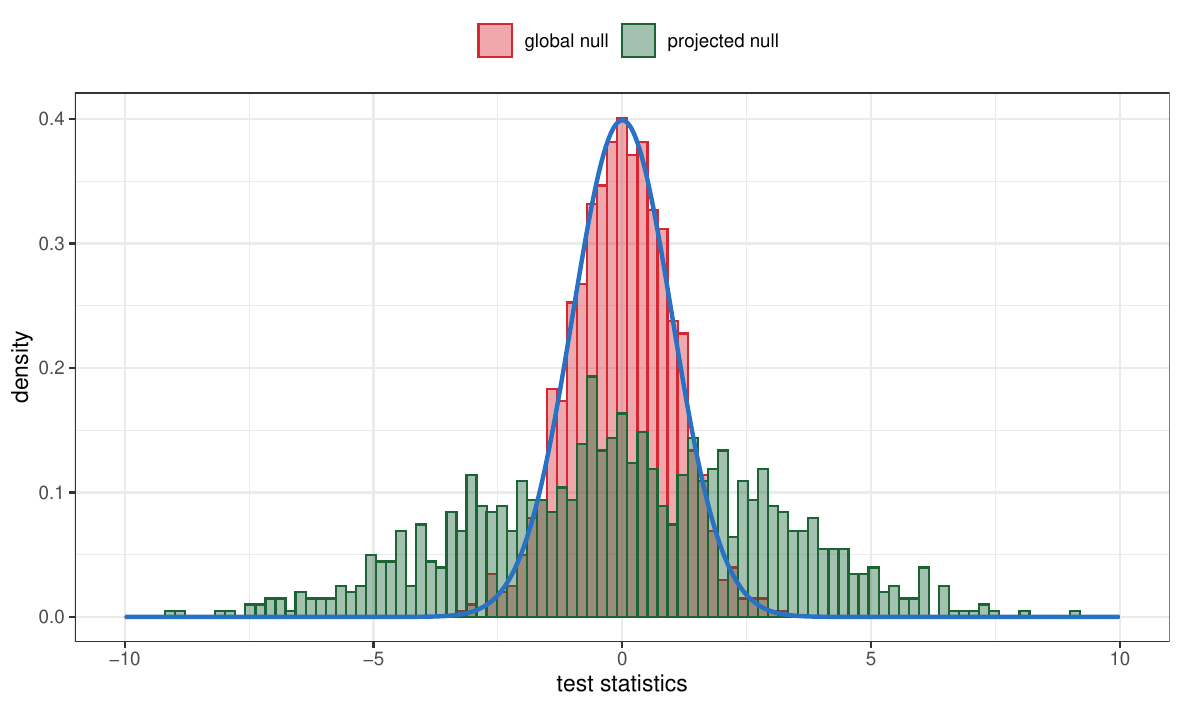}
    \caption{Histogram for the distribution of $T_{\rm pi}(v_1)$ under the global and projected nulls. The blue curve indicates the density of standard normal $\mathcal{N}(0,1)$. Sample size $N_X = 500, N_Z = 250$. Dimension $p = 100$.}
    \label{fig:cross-fitting under null}
\end{figure}

We illustrate the behavior of $T_{\rm pi}(u)$ in \Cref{fig:cross-fitting under null} with $u=v_1$.
We simulate the distribution of $T_{\rm pi}(u)$ under two settings and compare them with the standard normal density (the solid curve). In the first setting we have  $\mu_X = \mu_Z$, labeled as ``global null'' in the plot. This setting corresponds to a special point among all the distributions satisfying $(\mu_X - \mu_Z)^\top u = 0$. We can observe that the distribution of $T_{\rm pi}$ is close to a standard normal.
The second setting, the one labeled ``projected null'', corresponds to a more general case under the projected null, where the projected score $\left(\mu_X-\mu_Z\right)^{\top} u=0$ but $\mu_X \neq \mu_Z$. We observe that the distribution of $T_{\rm pi}$ is significantly over-dispersed compared to the standard normal. The lack of asymptotic normality of $T_{\rm pi}(u)$ under the null hypothesis makes it hard to determine a well calibrated rejection rule. 
The details of this simulation are listed in \Cref{app: for the figure}.  

\begin{remark}
    In the above definition of $T_{\rm pi}(u)$, we implicitly assume the eigenvector $u=v_1$ is well-defined, which requires a positive gap between the first and second largest eigenvalues of the covariance matrix. Corresponding plug-in $T_{\rm pi}(v_j)$ can also be considered for other eigenvectors $v_j$ so long as they are well-defined. Since the leading PC $v_1$ is identifiable only up to a sign. We assume the signs of the estimates are aligned such that $v_1^{(1)\top}  v_1^{(m)} \geq 0$  for all $m\in [M]$ when constructing \eqref{eq:simple statistics}. This convention also applies to the rest of the manuscript for discussion related to eigen-vectors and PCs. 
\end{remark}



\subsection{Intuition of One-step Correction}\label{section: intuition}

The intuition behind one-step estimation is the von Mises expansion, which is also seen as a distributional Taylor expansion in the semiparametric literature. In this subsection, we will use $v_1$ for the projection direction $u$.  We relate the testing problem to the (population) projected difference estimation:
\begin{equation}\label{eq:2.2-project-difference-parameter}
    \theta = (\mu_X - \mu_Z)^\top v_1\,.
\end{equation}
If one can establish an asymptotically normal estimator of $\theta$, then it can be directly applied to construct confidence intervals and derive a corresponding test for $\theta = 0$. 

To further simplify the presentation, we will consider a one-sample version of $\theta$ to illustrate the aforementioned principle. The extension to the corresponding two-sample problem is straightforward. Consider $\gamma = \mu_X^\top v_1$ and a plug-in estimator $\hat \gamma_{\rm pi}^{(-m)} = \mu_X^{(m) \top} v_1^{(-m)}$. One may attempt to use the following steps to characterize the asymptotic behavior of $\gamma_{\rm pi}^{(-m)}$:
\begin{equation}\label{eq: example decomposition}
      \gamma_{\rm pi}^{(-m)} - \gamma =  (\mu_X^{(m)} - \mu_X)^\top v_1 +  (\mu_X^{(m)} - \mu_X)^\top(v_1^{(-m)} - v_1)+ \mu_X^\top (v_1^{(-m)} - v_1).
\end{equation}
The first term in the RHS of \Cref{eq: example decomposition} converges to a normal distribution after proper normalization, which we will refer to as the CLT term. The second term is often of a higher order than the CLT term because both $\mu_X^{(m)} - \mu_X$ and $v_1^{(-m)} - v_1$ shrink to zero and are mutually independent. The third term corresponds to the bias due to using the estimated version $v_1^{(m)}$. When the dimension of $X$ is large, $v_1^{(-m)}$ often converges to $v_1$ at a slower rate than the CLT term and its irregular distribution (partially due to regularization such as sparsity induction) would dominate the CLT term. 

Fix $\mu_X$, we can treat $\mu_X^\top v_1^{(-m)}$ as a mapping from the distribution associated with $\mathcal{D}^{(-m)}$, denoted as $P_n^{(-m)}$, to a number. 
So the third term in \eqref{eq: example decomposition} can be rewritten as
\begin{equation*}
\mu_X^{\top}v_1^{(-m)}-\mu_X^{\top}v_1 =: f_\gamma(P_n^{(-m)}) - f_\gamma(P),
\end{equation*}
where we use $P$ to denote the law of $X$. The von Mises expansion states that for regular $f_{\gamma}$'s, we can perform the following Taylor expansion:
\begin{equation}\label{eq: example_von_mises}
   f_\gamma(P_n^{(-m)}) - f_\gamma(P) = -E_P[\varphi_\gamma(X;P_n^{(-m)})] + \text{higher order remainder }
\end{equation}
with a function $\varphi_\gamma$ that can sometimes be explicitly calculated---which is known as the influence function of the parameter $\gamma$. See \cite{fisher2021visually} for more intuition and visual illustrations on the expansion.

This motivates the bias correction procedure where people use a sample independent from $\mathcal{D}^{(-m)}$ to estimate the expectation wrapping $\varphi_\gamma$ in \eqref{eq: example_von_mises}, and add it on both sides of \eqref{eq: example decomposition}. Under certain conditions, we can show that 
\begin{equation*}
\gamma_{\mathrm{pi}}^{(-m)}-\gamma+E_{P_n^{(m)}}\left[\varphi_\gamma\left(X ; P_n^{(-m)}\right)\right]
\end{equation*}
is approximately normal, where $P_n^{(m)}$ denotes the empirical distribution given by $\mathcal D^{(m)}$. In addition to the CLT term in \eqref{eq: example decomposition}, the estimation error of $E_P\left[\varphi_\gamma\left(X ; P_n^{(-m)}\right)\right]$ also contributes to the final asymptotic distribution, and can be estimated using standard methods. 

Identifying the explicit form of the influence function is crucial to implement one-step correction. In our case, the relevant influence function estimators are \citep{magnus1985differentiating,critchley1985influence}
\begin{equation}\label{eq: definition nuisance}
\begin{aligned}
\phi_X^{(-m)}(X) & =s^{(-m)^{\top}}\left[\left(X-\mu_X^{(-m)}\right)\left(X-\mu_X^{(-m)}\right)^{\top}-\Sigma^{(-m)}\right] v_1^{(-m)}, \\
s^{(-m)} & =\left(\lambda_1^{(-m)} I_p-\Sigma^{(-m)}\right)^{+}\left(\mu_X^{(-m)}-\mu_Z^{(-m)}\right) .
\end{aligned}
\end{equation}
Here $\Sigma^{(-m)}$ is an estimate of the common covariance matrix $\Sigma = \Sigma_X = \Sigma_Z$ using samples in $\mathcal{D}^{(-m)}$, and $\lambda_1^{(-m)}$ is an estimate of $\lambda_1(\Sigma)$. 

\subsection{Debiased Tests for the Projected Null}\label{section: debiased test}
In this section, we develop a method for testing the projected null hypothesis $H_0^{\rm proj}(v_1)$ where $v_1$ is the top PC of the population covariance matrix. As we observed in \Cref{fig:cross-fitting under null}, the plug-in estimator combined with the given variance estimator does not approximate a standard normal. To address this issue, we leverage the one-step correction technique to achieve asymptotically normal test statistics. 

We propose using 
\begin{equation*}
T_{\mathrm{1s}}\left(v_1\right)=\hat{\sigma}_{\mathrm{1s}}^{-1} \sum_{m=1}^M \hat{\theta}_{\mathrm{1s}}^{(m)}
\end{equation*}
to test the projected null $H_0^{\rm proj}(v_1)$, where $\hat{\theta}_{\mathrm{1s}}^{(m)}$ is the one-step bias corrected projected difference estimator from a single fold split 
\begin{equation}\label{eq:debiased test statistics}
    \hat{\theta}_{\mathrm{1s}}^{(m)} = \hat{\theta}_{\rm pi}^{(m)}(v_1)+\left(n_X+n_Z\right)^{-1}\bigg\{\sum_{X_i \in \mathcal{D}^{(m)}} \phi_X^{(-m)}\left(X_i\right)+\sum_{Z_i \in \mathcal{D}^{(m)}} \phi_Z^{(-m)}\left(Z_i\right)\bigg\}\,.
\end{equation}

The influence function $\phi_X^{(-m)}$ was presented in \eqref{eq: definition nuisance}, and $\phi_Z^{(-m)}$ is similarly defined by replacing all the $X$ in $\phi_X^{(-m)}(X)$ by $Z$ but using exactly the same $s^{(-m)}$. Similar to $\hat \sigma_{\rm pi}$ in \eqref{eq: cross-fitting variance}, the quantity $\hat \sigma_{\rm 1s}$ is a sample-splitting estimate of the standard deviation of $\sum_{m=1}^M \hat{\theta}_{\rm 1s}^{(m)}$. We present its explicit formula in \Cref{app: explicit formula}. The parameter $\theta$ we analyzed is a two-sample version of $\gamma$ discussed in the previous section. The $\hat{\theta}_{\rm pi}^{(m)}(v_1)$ part in \Cref{eq:debiased test statistics} is the plug-in statistic and the rest is the one-step correction term.

Following the intuition we discussed in \Cref{section: intuition}, we can show that $\theta$ is a regular functional and the one-step correction will lead to a $\sqrt{n}$-consistent, asymptotically normal estimate.

\begin{assumption}\label{assumption: basic}
Let $\mathcal{D}$ be a collection of IID samples (more accurately, an IID triangular array detailed in \Cref{remark: triangular array}). The number of data splits $M$ is fixed. The leading eigenvalues of the covariance matrices are uniformly bounded:
\begin{equation*}
\lambda_1\left(\Sigma_X\right) \vee \lambda_1\left(\Sigma_Z\right) \leq C
\end{equation*}
for some constant $C>0$. The norm of mean difference, $\|\mu_X - \mu_Z\|$, is also bounded by a constant.
\end{assumption}

\begin{assumption}\label{assumption: degenerate}
There exists a sequence of vectors $u_n\in\mathbb{R}^{p_n}$, such that for each split $m\in[M]$ 
\begin{equation*}
\lim _{n \rightarrow \infty} E\left\|u^{(-m)}-u_n\right\|^2=0.
\end{equation*}
Define $W_n=\left[\left(X_{n 0}-\mu_{X, n}\right)^{\top} u_n\right]^2$, where $X_{n 0} \sim P_{X, n}$ is a random sample from a distribution that may vary with $n$. We assume the sequence $\left\{W_n\right\}$ is uniformly integrable:
$$
\lim _{t \rightarrow \infty} \sup _n E\left[W_n \cdot \mathbf{1}_{\{|W_n|>t\}}\right]=0.
$$
In addition, the projected variances are non-degenerate: $E\left[W_n\right] \geq C>0$ for some constant $C$. A similar condition holds for the $Z$ distribution.
\end{assumption}

\begin{remark}\label{remark: triangular array}(Varying dimensionality) In \Cref{assumption: degenerate}, we explicitly allow the dimensionality $p=p_n$ and the data-generating distributions to vary with the sample size $n$, a setup known as the triangular array setting in high-dimensional statistics. For each $n$, a dataset $\mathcal{D}$ of size $N_X+N_Z$ is drawn from distributions $P_{X, n}$ and $P_{Z, n}$, each supported on $\mathbb{R}^{p_n}$. As $n$ increases, new datasets are independently generated and reside in higher-dimensional spaces. The population quantities such as $\Sigma_X, \mu_X = \mu_{X,n}$ and $v_1$ also implicitly depend on $p_n$. The constants $C$ in \Cref{assumption: basic} and \ref{assumption: degenerate} serve as uniform constraints not depending on $n$. The uniform integrability conditions are used to establish a triangular array law of the large number for consistent variance estimation. For simplicity, we will suppress most subscripts when their meaning is clear from context.
\end{remark}

Our main result in this section is formally presented below. We use $\phi_{X,n}$ to denote the ``true" influence function, replacing all the estimated quantities in $\phi_X^{(-m)}$ by their population version. See \eqref{eq: define_true_influence} for its explicit formula. While $\phi_{X, n}$ depends on $n$ in the triangular array setting, we will suppress the index $n$ when the meaning is clear from context. We will also use $w$ to denote the training sample size ratio $w=n_X /\left(n_X+n_Z\right)$.

\begin{theorem}
\label{th: one step}
    Suppose $T_{\rm 1s}(v_1)$ is calculated from a sample satisfying \Cref{assumption: basic} and \Cref{assumption: degenerate} holds for $u = v_1$. We further assume 
    \begin{itemize}
         \item Non-vanishing variance: $0 < \operatorname{Var}\left\{\left(X-\mu_X\right)^{\top} v_1+w \phi_X(X)\right\}$. A similar condition holds for $Z$.
          \item Identical covariance matrices: $\Sigma_X = \Sigma_Z$ and $\lambda_1(\Sigma_X) - \lambda_2(\Sigma_X) > 0$.
          \item The squared influence functions $\phi_X^2(X) = \phi^2_{X,n}(X_{n0})$ are uniformly integrable (\Cref{remark: triangular array}). 
          Its estimates converge in $L_2$: \begin{equation}\label{eq: nuisance moment convergence}\lim_{n\rightarrow\infty} E[(\phi^{(-m)}_X(X) - \phi_X(X))^2] = 0.
          \end{equation}
          A similar condition also holds for $\phi^{(-m)}_Z$. 
         \item The nuisance parameters are estimated well: for any $\epsilon > 0$:
         \begin{equation}\label{eq: quarter rate}
             \lim_{n\rightarrow\infty}\mathbb{P}\left(\left\|\Sigma^{(-m)}-\Sigma\right\| \geq \epsilon n^{-1 / 4}\right) = 0.
         \end{equation}
         Similar conditions also hold for $\left\|\mu_X-\mu_X^{(-m)}\right\| $ and $\left\|\mu_Z-\mu_Z^{(-m)}\right\|$.
    \end{itemize}
Then under the projected null hypothesis $H_0^{\rm proj}(v_1)$, we have $T_{\rm 1s}(v_1) \stackrel{d}{\rightarrow} \mathcal{N}(0,1)$ as $n\rightarrow\infty$. 
\end{theorem}

The proof of \Cref{th: one step}
is presented in \Cref{app: debiased stat}.

Condition \eqref{eq: quarter rate} in \Cref{th: one step} is the most essential for establishing the asymptotic normality. They require the high-dimensional quantities $\Sigma$, $\mu_X,\mu_Z$ to be estimated at a rate faster than $n^{-1/4}$ (recall in low-dimensional settings they can be estimated in a parametric rate $\sqrt{p/n}$). This type of condition is common in the one-step estimation literature---including the well-known doubly-robust estimator of average treatment effect \citep{glynn2010introduction}. When $p = p_n$ diverges faster than $n^{1/2}$, the above $\sqrt{p/n}$ rate no longer satisfies our requirement. Additional structures and regularization techniques are necessary to improve estimation accuracy.

In our case, we can apply some regularized estimators of $\mu_X, \mu_Z$ to achieve the $o_P(n^{-1/4})$ rate. One choice is simply calculating simple sample means from $\mathcal{D}^{(-m)}$ and hard threshold each entry at $\sqrt{\log p/n}$. This procedure and a close variation (``soft-thresholding") give estimators converging in rate $\sqrt{\log p/n} \ll n^{-1/4}$, assuming a small number of entries of $\mu_X$ are non-zero \cite{johnstone2019gaussian}. In the statistical literature, this type of estimator has been extensively discussed in wavelet nonparametric regression in the 1990s \cite{donoho1995noising}. It is also related to James-Stein estimator \cite{stein1956inadmissibility} and Lasso under orthonormal designs (\cite{tibshirani1996regression}, Section~10).

Estimation of high-dimensional covariance matrices is a more recent topic and has been extensively studied in the past two decades. The high dimensionality is often tackled by some covariance structures such as low-rank, approximate block-diagonal, or sparsity. The theoretical rates of many estimators, measured in the operator spectral norm $\|\Sigma^{(-m)} - \Sigma\|$, are often of order $\sqrt{\log p/n}$ or $n^{-\alpha/(2\alpha + 1)}$ with some regularity index $\alpha > 0$, possibly achieving the required $o(n^{-1/4})$ rate in \eqref{eq: quarter rate}. We refer our readers to \cite{fan2016overview, cai2016review, lam2020high} for more extensive surveys of frequently imposed structures and available methods.

\begin{remark}
    Condition \eqref{eq: quarter rate} may imply condition \eqref{eq: nuisance moment convergence} under certain boundedness conditions on the components of $\phi_X,\phi_Z$ (convergence in probability does not unconditionally imply convergence in moments). Since they are neither sufficient nor necessary for each other and control different elements in the proof, we state them separately. For semi-parametric estimation without sample-splitting, condition \eqref{eq: nuisance moment convergence} needs to be modified to a stronger version restricting the estimates in a Donsker class (e.g. \cite{kennedy2022semiparametric} Section 4.2).
\end{remark}

\section{Approximate Orthogonality and Anchored Projection }\label{section: app orthogonality}

The previous section presents a prototypical one-step inference procedure for valid inference of the general projected mean difference parameter in \eqref{eq:2.2-project-difference-parameter}. Our theoretical and numerical study reveals that, interestingly, the one-step bias correction is not always necessary for asymptotic normality, and the plug-in statistic $T_{\rm pi}$ can be asymptotically normal under certain conditions. In this section, we characterize one sufficient condition for asymptotic normality of $T_{\rm pi}$. These results are particularly useful when influence functions are unknown. Specifically, it allows $u^{(-m)}$ to be calculated from black box algorithms whose explicit expression is less explicit. Building on top of this result, we develop an ``anchored projection'' test that enjoys better power against the global null hypothesis without suffering from the degeneracy issue.

\subsection{Approximate Orthogonality}\label{subsec:3.1-approx-orth}
One simple scenario for asymptotic normality of $T_{\rm pi}$ is under the global null $\mu_X = \mu_Z$, as shown in \Cref{fig:cross-fitting under null}.  More generally, a sufficient condition is the ``approximate orthogonality'' \eqref{eq:3.2-approx-orth} in the following theorem.

\begin{theorem}\label{th: approximate orthogonal}
    Assume \Cref{assumption: basic} and \Cref{assumption: degenerate} hold. If 
    \begin{equation}
        \label{eq:3.2-approx-orth}
        (\mu_X-\mu_Z)^T u^{(-m)} = o_P(n^{-1/2})
    \end{equation}
    for all $m\in[M]$, then
     $T_{\mathrm{pi}}(u) \xrightarrow{d} \mathcal{N}(0,1)$ as $n \rightarrow \infty$.
\end{theorem}

The proof of \Cref{th: approximate orthogonal} is presented in \Cref{app: app_orthogonal}.
A useful special case of approximate orthogonality is when there exists a subset $\mathcal S\subseteq[p]$ such that
\begin{equation}\label{eq: partial mean identical}
        \mu_{X,j} = \mu_{Z,j},\quad\text{for all 
 }j\in \mathcal{S}
    \end{equation} and for all $m\in[M]$:
    \begin{equation}\label{eq: u has small residual}
        \sum_{j \notin \mathcal{S}}\left(\mu_{X,j}-\mu_{Z,j}\right) \cdot u^{(-m)}_j=o_P(n^{-1/2}),
    \end{equation}
where the subscript $j$ in \eqref{eq: partial mean identical} and \eqref{eq: u has small residual} corresponds to the $j$-th element of a $p$-dimensional vector.  As a further special case, under the global null ($\mu_X=\mu_Z$), both \eqref{eq: partial mean identical} and \eqref{eq: u has small residual} are directly satisfied with $\mathcal S=[p]$. In general, we only need $u^{(-m)}$ to be approximately orthogonal to $\mu_X - \mu_Z$ on the complement 
 of signal dimensions $\mathcal{S}^c$ as stated in \Cref{eq: u has small residual}. In particular, any $u^{(-m)}$ with ${\rm supp}(u^{(-m)}) \subset \mathcal{S}$ satisfies this condition. When we observe a large $T_{\rm pi}$ calculated from such a sparse $u^{(-m)}$, we should expect there are some dimensions within $\mathcal{S}$ to have mean shifts.

\begin{figure}[!t]
    \centering
\includegraphics[width =0.7\linewidth]{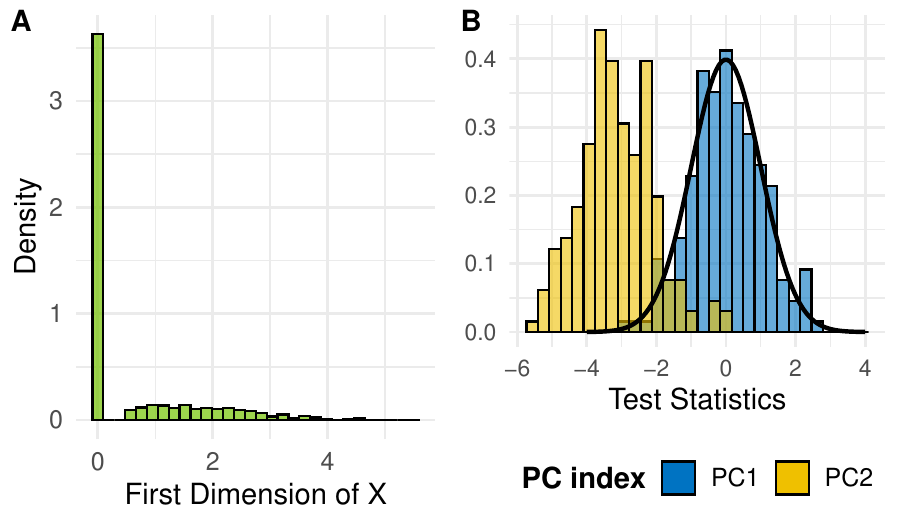}
    \caption{A low $n,p$ ratio experiment. \textbf{(A)} The distribution of original data is highly zero-inflated. \textbf{(B)} Histogram of $T_{\rm pi}(v_j), j = 1,2$. PC1 is close to the standard Gaussian indicated by the black line; PC2 captures the mean difference signal. Simulation detail can be found in \Cref{app: simulation truncated normal}.}
    \label{fig: marginal and plug in}
\end{figure}

We set up a simulation study to illustrate an application scenario of \Cref{th: approximate orthogonal} and examine the promised Gaussianity. We generated two $p=300$ independent samples with $N_X = 250$ and $N_Z = 50$. The covariance structure is block-diagonal with block size $=10$ (so $30$ blocks in total). We denote them as $\mathcal{S}_i = \{10(i-1)+1,...,10i\}$. The entries of the samples are zero-inflated Gaussian, and we plotted the distribution of the first dimension of $X$ in \Cref{fig: marginal and plug in}A. The marginal distribution is designed to be close to normalized scRNA sequencing data with a significant portion taking exactly $0$ (in our case $\sim 65\%$ are zero). The $n,p$ ratio is also close to many real scRNA sequencing datasets.  

The mean difference between $\mu_X-\mu_Z$ is on the second block: ${\rm supp}(\mu_X-\mu_Z) = \mathcal{S}_2$. The true leading PCs satisfy ${\rm supp}(v_j) = \mathcal{S}_j$ for $j = 1,2$. We apply sPCA to estimate the leading PCs. The PC1, $v_1^{(-m)}$, has larger non-zero loadings in the first block $\mathcal{S}_1$ and has some non-zero values in the other blocks due to randomness. On the other hand, $v_2^{(-m)}$ are mostly supported on the second block $\mathcal{S}_2$ where the mean difference signal is established. Thus $v_1^{(-m)}$ satisfies the condition in Theorem \ref{th: approximate orthogonal} while $v_2^{(-m)}$ does not. The distribution of $T_{\rm pi}(v_1)$ and $T_{\rm pi}(v_2)$ are shown in \Cref{fig: marginal and plug in}B. Although the marginal distributions are irregular, it is not hard to obtain an approximately Gaussian statistic under the given moderate sample sizes. We note that sPCA is crucial for this experiment---using least square PCA leads to inflated type-I error because they do not localize ${\rm supp}(v_j^{(-m)})$ and assign loadings to all the dimensions, violating \Cref{eq: u has small residual}.

In addition to sPCA, many other methods, including the popular clustering tool WGCNA \cite{langfelder2008wgcna}, can be applied to obtain the projection direction $u^{(-m)}$. Based on the correlation structure between the dimensions, WGCNA divides the total $p$ dimensions into multiple modules $\hat{\mathcal{S}}_i\subset [p]$. Within each module, it would perform PCA to obtain a vector $u_i^{(-m)}$ supported on $\hat{\mathcal{S}}_i$, which is called eigengene (\cite{langfelder2007eigengene}, equation (29)), serving as a summary of the variation pattern. \Cref{th: approximate orthogonal} can also be applied to this setting.

\paragraph{Comparison of $T_{\rm pi}$ and $T_{\rm 1s}$}
The approximate orthogonality condition \eqref{eq:3.2-approx-orth} is more likely to hold under null hypotheses, such as the global null $\mu_X=\mu_Z$.   Thus, a small p-value is obtained from $T_{\rm pi}(u)$ usually provides evidence against the global null hypothesis, but does not provide much information about the alternative. In contrast, if $T_{\rm 1s}(u)$ returns a small p-value, we not only know a difference likely exists within the support of $u$, but also obtain a valid confidence interval of the projected difference $(\mu_X - \mu_Z)^\top u$ through the debiased estimate.

\subsection{Anchored Projection Tests}
\label{section: anchored lasso}

In this previous subsection we showed that, according to the approximate orthogonality condition, the plug-in projected statistic $T_{\rm pi}$ can serve as a convenient and flexible tool to detect deviation from the global null hypothesis $\mu_X=\mu_Z$. In this section, we formally develop this idea into an ``anchored projected test'' with the following appealing features:
\begin{itemize}
    \item It does not involve debiasing; 
    \item It achieves good power against the global null;
    \item It avoids the degeneracy issue commonly encountered by existing methods under the null hypothesis;
    \item It provides information on the potential location of the signal under sparse alternatives.
\end{itemize}

Although the sparse PCs serve as a natural and reasonable choice of projected comparison,
depending on the scientific research goal, one may alternatively be interested in projective directions that maximize the contrast between the two groups, prioritizing overall detection power over the correlation structure. Intuitively, these directions would correspond to the linear discriminating directions that best classify the two populations.

Constructing high-dimensional sparse linear classifiers has been well-studied in the literature, including logistic Lasso \cite{tibshirani1996regression,van2008high} and sparse LDA \cite{shao2011LDA, cai2011direct}. However, when $\mu_X = \mu_Z$, the population-level discriminating direction degenerates. In practice, it is also direct to verify via a simple simulated experiment (\Cref{appfig: funny lasso}) that cross-validated linear classifiers such as logistic Lasso have a positive probability to be exactly zero. This is a common problem encountered in two-sample testing involving nuisance parameters \cite{liu2022multiple,williamson2023general,dai2022significance,lundborg2022projected}, and many existing results \cite{liu2022multiple} are only established under the alternative hypothesis. 

In order to overcome the degeneracy issue, we propose an easy-to-use sparse projection that ``anchors" the potentially degenerative discriminating direction to a regular proxy such as a sparse PC vector. When the signal is moderately strong, the projection direction will mainly follow the estimated discriminating direction, which better contrasts the samples and yields higher power. On the other hand, when the signal is weak, the estimated discriminating direction is noisy, and the proxy direction takes over to avoid degeneracy.

Let $\beta^{(-m)}$ be a discriminating direction estimated from $\mathcal{D}^{(-m)}$, using each sample in $\mathcal{D}^{(-m)}$ as the covariate and group label (control or treatment) as the response. The proposed anchored projection test statistic takes the following form:

\begin{equation}
\label{eq:Lasso testing statistics}
    T_{\rm anc}(v,\beta) := \hat \sigma_{\rm anc}^{-1}\sum_{m=1}^M \left(\mu_X^{(m)}-\mu_Z^{(m)}\right)^{\top} \left(v^{(-m)}+w_n \beta^{(-m)}\right).
\end{equation}
The normalizing standard error $\hat \sigma_{\rm anc}$ is similarly defined as $\hat \sigma_{\rm pi}$ in \eqref{eq: cross-fitting variance}, replacing $ u^{(-m)}$ by the hybrid projection vector $  v^{(-m)}+w_n \beta^{(-m)}$.
The weight parameter $w_n\in\mathbb{R}$ diverges as $n \rightarrow \infty$ is a hyperparameter of the method, which shifts the projection direction towards $\beta^{(-m)}$ when the signal is strong. Under $\mu_X = \mu_Z$, the $v^{(-m)}$ component dominates so long as $w_n$ does not diverge too fast, avoiding degeneracy and allowing for tractable distribution of $T_{\rm anc}$. The choice of discriminating direction estimate $\beta^{(-m)}$ can be quite flexible.
We have the following distributional guarantee.


\begin{corollary}
\label{corollary: anchored Lasso}
     Under \Cref{assumption: basic} and assuming \Cref{assumption: degenerate} holds for $v^{(-m)}$, we further require that
\begin{equation}\label{eq: classifier_quality}
         \lim_{n\rightarrow\infty}E\ \left\|w_n \beta^{(-m)}\right\|^2 = 0.
     \end{equation}
      Then when $\mu_X = \mu_Z$ we have $T_{\rm anc} \stackrel{d}{\longrightarrow} \mathcal{N}(0,1)$ as $n \rightarrow \infty$.
\end{corollary}

\begin{proof}[Proof of \Cref{corollary: anchored Lasso}]
By \Cref{assumption: degenerate}, we know there is a sequence of $v_n$ that $v^{(-m)}$ converges to. Denote $u^{(-m)} = v^{(-m)}+w_n \beta^{(-m)}$, we then have,
\begin{equation*}
    E\|u^{(-m)} - v_n\|^2 \leq 2E\|v^{(-m)}\|^2 + 2E\|w_n \beta^{(-m)}\|^2 \rightarrow 0.
\end{equation*}
So we know \Cref{assumption: degenerate} also holds for $u^{(-m)}$. Moreover, conditions \eqref{eq: partial mean identical} and \eqref{eq: u has small residual} are satisfied under $\mu_X = \mu_Z$. Now the result follows directly from \Cref{th: approximate orthogonal}.
\end{proof}

\begin{remark}(Power of the anchored test)
The discriminating direction $\beta$ can be related to the distributions of $X$ and $Z$ through a classification problem. We associate each sample point in the pooled data $\mathcal{D}$ a binary label  $Y$, depending on whether this sample comes from the $X$ or $Z$ population. We denote the best linear discriminating direction (or the logistic regression coefficient) as $\beta$, which can be estimated using the corresponding high-dimensional sparse estimators \citep{van2008high,cai2011direct}.
Under mild assumptions, $\mu_X \neq \mu_Z$ implies a non-zero $\beta$. 
Therefore, the test based on the anchored projection statistic $T_{\rm anc}$ has power converging to $1$, so long as $\|\beta^{(-m)}-\beta\|=o_P(\|\beta\|)$ and $w_n\|\beta\|\rightarrow \infty$.
\end{remark}

In practice, we also found a thresholded-version of $\beta^{(-m)}$ works as well:
\begin{equation*}
u^{(-m)} = v^{(-m)} + w_n \beta^{(-m)} \cdot \mathbf{1}\{ \| \beta^{(-m)} \| \geq r_n \}
\end{equation*}
for some threshold level $r_n \geq 0$. This allows us to use a large $w_n$ so that $u^{(-m)}$ aligns better with $\beta^{(-m)}$ when the signal surpasses the threshold.
    
The theoretical choice of the threshold $r_n$ depends on the rate of convergence of the original estimate $\beta^{(-m)}$. When the true regression coefficient $\beta$ is zero, in typical high-dimensional sparse classification settings we usually have $\|\beta^{(-m)} - \beta\| = \|\beta^{(-m)}\| = O_P(n^{-1/2}\sqrt{\log p})$, so that the anchoring test statistic will offer asymptotically valid null distribution as long as $\lim_{n\rightarrow\infty} r_n/(\log p/n)^{1/2} = \infty$. 
In our numerical examples, the choice of $r_n = n^{-1/3}$ has worked reasonably well.  With this $r_n$, the choice of $w_n$ becomes less sensitive, and we use $w_n = \sqrt{n}$ in both simulation and real-data analysis. We will proceed with this choice of $u^{(-m)}$ in \Cref{section: simulation} \& \ref{section: real data}---Logistic Lasso estimates $\beta^{(-m)}$ and sPCA proxy $v_1^{(-m)}$. If one replaces the \Cref{eq: classifier_quality} with $\lim _{n \rightarrow \infty} \mathbb{P}\left(\left\|\beta^{(-m)}\right\| \geq r_n\right)=0$, the related $T_{\rm anc}$ is also asymptotically normal under $\mu_X = \mu_Z$, using a similar argument as \Cref{corollary: anchored Lasso}. 

When $r_n = 0$, the shrinkage function reduces to an identity mapping \eqref{eq:Lasso testing statistics}. In this case we can usually take $w_n = n^{\alpha}$ for some $\alpha\in (0,1/2)$. In \Cref{section: clearys data}, we adopt this setting, using a combination of Logistic group-Lasso estimates for $\beta^{(-m)}$ and sPCA $v_1^{(-m)}$.



\section{Simulation Studies}\label{section: simulation}

In this section, we present some numerical results based on simulated datasets. We are interested in the performance of $T_{\rm pi}$, $T_{\rm 1s}$ and $T_{\rm anc}$ as well as a literature method for comparison \cite{chen2010two}. The existing method is a popular, powerful procedure for testing $\mu_X = \mu_Z$ and is more favored over other existing methods when there are small signals in most dimensions (the $L_2$-type alternative in \cite{huang2022overview}). The authors also applied their method to some gene-set comparison problems. 

We simulate data under three scenarios: the global null $\mu_X = \mu_Z$; a strictly weaker projected null $H_0^{\rm proj}(v_1)$, with $\mu_X\neq \mu_Z$ but $(\mu_X - \mu_Z)\perp v_1$; and the alternative hypothesis $(\mu_X - \mu_Z)^\top v_j \neq 0$, $j = 1,2$. That is, in the alternative hypothesis setting, there are signals aligning with both population PC1 and PC2. In this section, we will focus on the validity and power of the tests. The interpretation aspect will be explored in the real-data example. We reject the null hypothesis when the absolute value of the test statistic is greater than $97.5\%$-quantile of $N(0,1)$. When $\mu_X = \mu_Z$, we expect the three discussed statistics to have an approximate $0.05$ rejection proportion. For $H_0^{\rm proj}(v_1)$, only $T_{\rm 1s}$ is expected to have a $0.05$-size, while the other two should have a larger size. Under the alternative hypothesis, we prefer a test that rejects more often implying a better power.

We consider a zero-inflated normal distribution of $P_X$ and $P_Z$. The sample matrix would have a significant proportion of exact zeros, mimicking normalized scRNA data where gene expression reads are highly sparse. We use $N_X = N_Z\in \{100, 300, 500\}$. Sample dimension $p = 10^3$. The samples have a sparse, spiked covariance structure \cite{johnstone2001distribution}. See \Cref{app: simulated data} for a complete description of simulation details.

\begin{figure}[!t]
    \centering
    \includegraphics[width =0.9\linewidth]{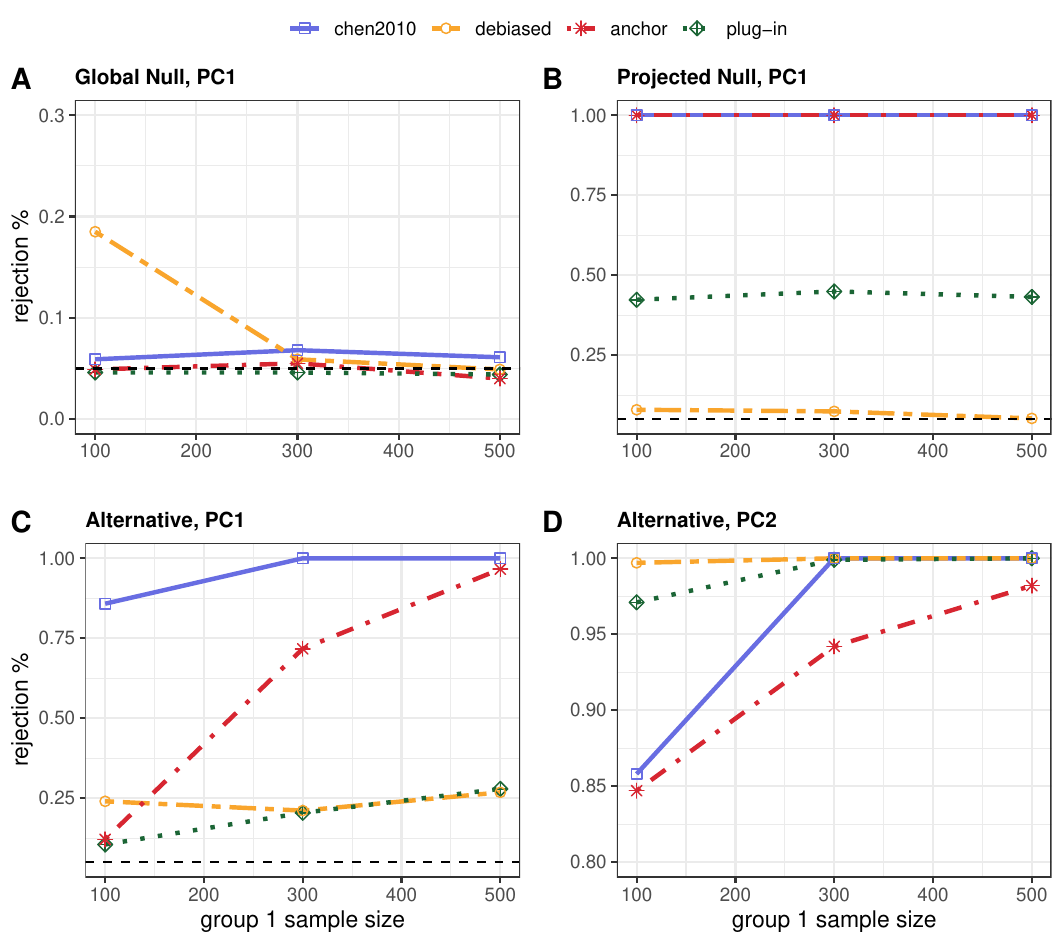}
    \caption{Significant results of the numerical studies. The title of each subplot should be read as: (simulation setting), (PC used for estimator construction). 
    {\tt chen2010} the literature method, {\tt debiased}, $T_{\rm 1s}(v_1)$ in A-C, $T_{\rm 1s}(v_2)$ in D; {\tt anchor}, $T_{\rm anc}(v_1)$ in A-C, $T_{\rm anc}(v_2)$ in D; {\tt plug-in}, $T_{\rm pi}(v_1)$ in A-C, $T_{\rm pi}(v_2)$ in D.} 
    \label{fig: main simulation}
\end{figure}

The rejection proportion of each test in different settings is estimated with $10^3$ Monte Carlo repeats, and the results are presented in \Cref{fig: main simulation}. Under the global null hypothesis when there is absolutely no signal (\Cref{fig: main simulation}, A), all of the methods have well-calibrated rejection proportions when sample sizes are greater than $300$. The debiased test statistics $T_{\rm 1s}$ has an inflated type I error when the sample size is small. 

Under the projected null, $T_{\rm 1s}$ meets the expected $0.05$ rejection proportion with larger sample sizes (\Cref{fig: main simulation}, B). Although the difference is orthogonal to $v_1$, the absolute norm of the difference $\|\mu_X - \mu_Z\|$ is set to be large, which makes $T_{\rm anc}$ and the literature method always reject. The plug-in statistic $T_{\rm pi}$ also shows some ``power'', but this implies it cannot be used as a valid test for $H_0^{\rm proj}(v_1)$ although it is tempting to apply it to this case. 

The results in \Cref{fig: main simulation}, C \& D correspond to the same simulation setting (alternative hypothesis), but the methods under comparison are different. We consider $T_{\rm pi}(v_1)$, $T_{\rm 1s}(v_1)$ and $T_{\rm anc}(v_1)$ that target/anchor at PC1 in subplot C, whereas in panel D it is their PC2-version being assessed. The literature method ${\tt chen2010}$ is identical across the two subplots. The signal aligned with PC1 $v_1$ is set to be smaller than that with PC2 $v_2$, therefore the observed rejection rate is, in general, lower in panel C than D. The ${\tt chen2010}$ method can leverage the signal from both $v_1$ and $v_2$ and appear to be more powerful than the PC1 versions (but less than PC2 versions). Notably, the anchored-test $T_{\rm anc}(v_1)$ can adaptively adjust the projection direction to where the stronger signal lies, even when it is anchored to the sub-optimal direction $v_1$ (\Cref{fig: main simulation}~C). 

\section{An Application using Perturb-seq Data}\label{section: clearys data}

\subsection{Dataset and Pre-processing}

To interrogate the function of 598 immune-related genes,  \citet{yao2024scalable} employed a functional genomic approach called Perturb-seq \cite{Dixit2016,Replogle2022,Schraivogel2020}.  For each cell, one of the targeted immune-related genes was perturbed (knockout) using CRISPR-Cas9, and then all cells were manipulated to trigger a strong immune response. In total, the experimental setup involved 599 groups of cells, each group had a specific gene knockout, except for one control group, which had no treatment. Single cell RNA sequencing was applied to assess the resulting gene expression changes across the whole genome in response to each gene knockout. The scientific objective was to analyze and compare the transcriptional profiles between these cells, thereby gaining insight into the molecular underpinnings of genes associated with the immune response.  

Among the 599 perturbed groups, which exhibited varying sample sizes, we focused our analysis on a subset of groups with higher cell counts. Specifically, we included all 50 perturbations presented in the original publication \cite[Figure 3D, left]{yao2024scalable}, where the sample sizes of the analyzed groups ranged from 41 to 173 cells. The control group contained a large number of cells (4492), and to accelerate computation, we randomly subsampled 500 control cells for comparison. 

We then regressed out the impact of cell-cycle phase and library size (detailed in \Cref{app: cleary_pre_processing}). All $p$ gene expression features were further normalized to have sample variance equal to 1. To perform the comparison, we applied group-lasso as the classifier and calculated $T_{\rm anc}$ with
\begin{equation*}
    u^{(-m)} = v_1^{(-m)} + n^{1/3} \beta^{(-m)}_{\rm GLasso}.
\end{equation*} 

The gene module information, used as input for the group-lasso, was established using the control cells only. Our pipeline incorporated correlation structures identified by \texttt{CSCORE} \cite{su2023cell}, \texttt{WGCNA} \cite{langfelder2008wgcna}, and Gene Ontology (GO) criteria, as described in \Cref{app: cleary_pre_processing}. Leading PC $v_1^{(-m)}$ is estimated using sPCA \texttt{PMA} in package \cite{PMA}. We use $M=5$ when performing cross-fitting.

\subsection{Test Results}

\begin{figure}[!t]
    \centering
    \includegraphics[width =\linewidth]{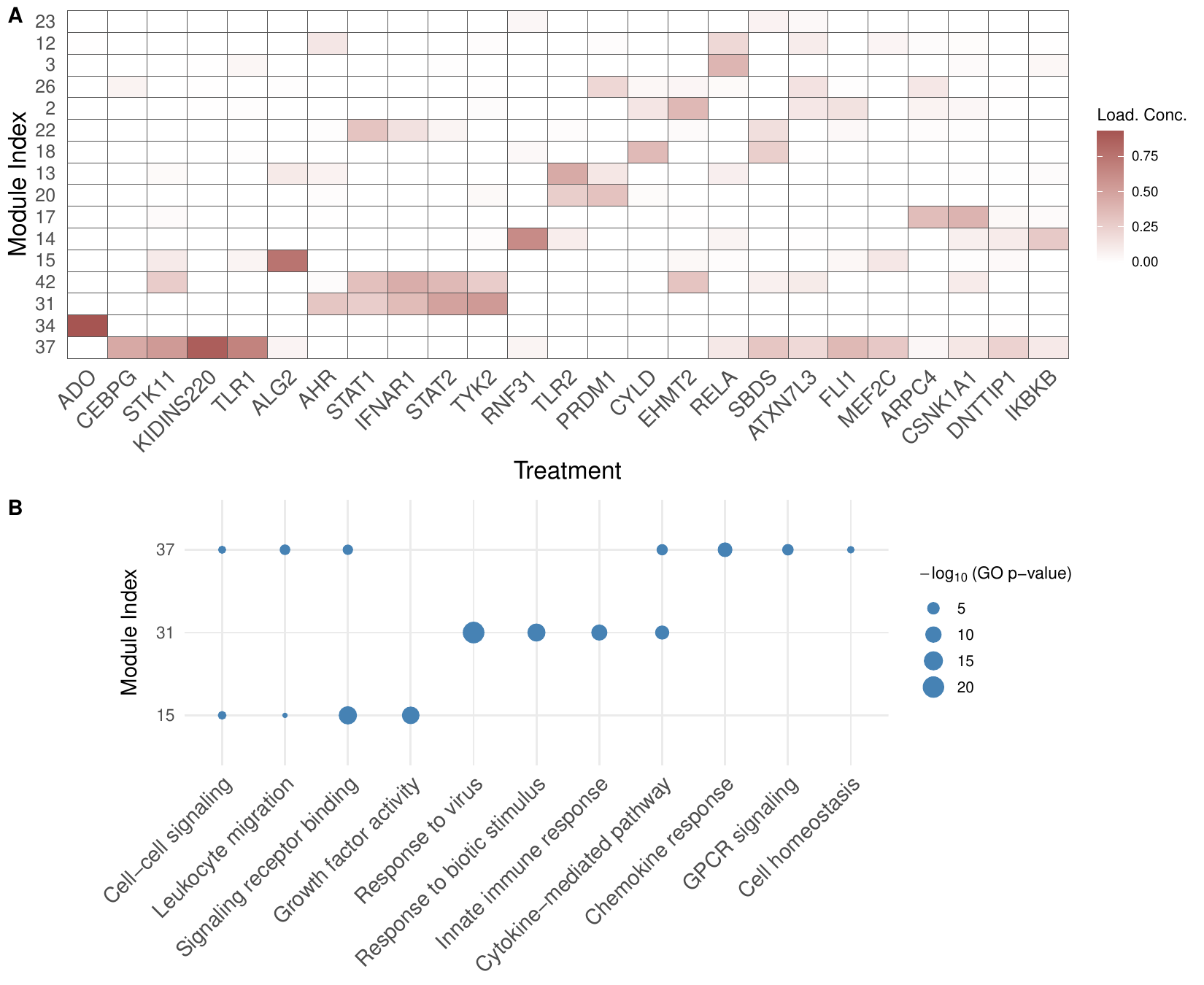}
    \caption{Perturb-seq data analysis results. 
    \textbf{A}. Loading concentration of the projection direction for selected significant perturbations and gene modules. Darker color implies more contribution of gene groups to the projection direction $\bar u$. 
    \textbf{B}. GO enrichment analysis for three selected gene modules. The corresponding GO IDs for each pathway are provided in \Cref{tab:short_desc_go_ids}.}
    \label{fig: cleary_loading_concentration}
\end{figure}

In \Cref{fig: cleary_loading_concentration}A, we present a subset of perturbation groups whose p-values associated with $T_{\rm anc}$ are smaller than 0.05. We use standard normal as the reference distribution according to \Cref{corollary: anchored Lasso}. Coloring intensity reflects the loading concentration of selected gene modules, defined as follows. For each perturbation-control pair, we compute the average projection direction $\bar u = M^{-1}\sum_{m = 1}^M u^{(-m)}$. For each of the 42 gene modules $\mathcal{S}_g \subset [p]$, $g\in [42]$, the loading concentration is $\sum_{i\in\mathcal{S}_g} \bar u_i ^2 / \|\bar u\|^2$. A higher loading indicates a greater contribution of the gene module to the projection direction and test result. The loading concentration of $\bar u$ is mostly driven by $\beta_{\rm GLasso}^{(-m)}$ due to the diverging weight $w_n = n^{1/3}$.

For the perturbation ADO, the majority of the discriminative power (of the group-Lasso classifier) is attributed to genes in module 34. A similar pattern is observed for ALG2 and RNF31, each exhibiting active groups that are uniquely associated with them.

We also observe overlap in the projective structure between certain perturbations. Notably, the projective directions for CEBPG, STK11, KIDINS220, and TLR1 are all concentrated within module 37, implying potential functional similarity. Likewise, AHR, STAT1, IFNAR1, STAT2, and TYK2 induce a common impact on modules 31 and 42.

Gene Ontology is a framework facilitating gene function description. A GO enrichment analysis provides p-values that quantify how significantly a set of genes is overrepresented in a specific functional category, compared to what would be expected by chance. In \Cref{fig: cleary_loading_concentration}B, we present the GO enrichment results for selected gene modules. The modules are primarily enriched in pathways related to immune response and cell signaling. Notably, the functional annotations of Module 15 are almost a subset of those of Module 37. The pathways enriched in Module 31 show less overlap with those in Modules 15 and 37. It is scientifically intriguing to investigate the functional similarity among AHR, STAT1, IFNAR1, STAT2, and TYK2, all of which have unique impacts on this module. Genes in Module 34 do not show significant enrichment for any pathway after Bonferroni correction. Module 42 consists mainly of mitochondrial genes. We retain Module 42 in the analysis due to its interesting overlapping pattern with Module 31 in \Cref{fig: cleary_loading_concentration}A.

\section{An Application to Immune Cell Gene Expression in a Lupus Study}\label{section: real data}

\subsection{Dataset and Pre-processing}


To investigate molecular mechanisms underlying Systemic Lupus Erythematosus (SLE)—a heterogeneous autoimmune disease with elevated prevalence in women and individuals of Asian, African, and Hispanic ancestry—we apply our proposed procedures, $T_{\rm 1s}$ and $T_{\rm anc}$, to a large-scale single-cell RNA-sequencing study \cite{perez2022single}. One of the study’s primary goals is to identify differentially expressed genes across immune cell types between SLE cases and healthy controls. The public dataset contains expression profiles of $1.2$ million cells from 8 major immune cell types, sampled from $261$ individuals ($162$ with SLE and $99$ healthy controls).

We use the Python package {\tt scanpy} \cite{wolf2018scanpy} to pre-process the single-cell data and select the top $2000$ highly variable genes within each cell type.  For each cell type, we aggregate expression across cells from the same individual to obtain ``pseudo-bulk" counts for each gene, and then remove genes expressed in less than 10 individuals.  This means each sample of our analysis corresponds to one individual and they can be treated as IID samples from several homogeneous populations.  Next, we applied the standard shifted-log-normalization with a size factor (e.g., equation (2) in \cite{ahlmann2023comparison}), converting raw expression count to its logarithm, to stabilize the sample value and make it more amenable to comparisons. In this study, we focus on $4$ important immune cell types with a moderately large number of samples and compare the case and control gene-expression profiling within each. We also regress out the library size, sex, population, and processing cohorts to remove potential confounding under a simple linear model. The gene expression variable (i.e. each dimension of $X_i, Z_i$) is normalized to have unit variance. We use $M=10$ when performing cross-fitting.

\begin{figure}[!t]
    \centering
    \includegraphics[width =0.8\linewidth]{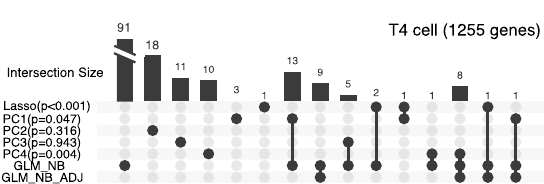}
    \caption{Test results for the T4 cell sample. We report the p-value for each proposed test. Lasso: $T_{\rm anc}\left(v_1\right)$; PC1-PC4: $T_{\rm 1s}\left(v_1\right)$ - $T_{\rm 1s}\left(v_4\right)$. We also assess how much the active gene set--those having non-zero loadings in Lasso or PC vectors--of each test overlaps, in the format of an ``UpSet'' plot. Two lists of genes that are reported to be marginally significant between groups are also included. Specifically, {\tt GLM\_NB} is based on negative-binomial regression with (threshold $0.05$) and {\tt GLM\_NB\_ADJ} has a Bonferroni adjusted threshold $0.05/1255$. This plot should be read as: there are $91$ genes reported to be significant according to negative-binomial regression but not contained in any of Lasso or PC vectors; There is one gene contained in both Lasso and PC1 active sets.}
    \label{fig: upset}
\end{figure}

\subsection{Testing Results}

The anchored-Lasso test $T_{\rm anc}\left(v_1\right)$ and the debiased test for the top four PC directions $T_{\rm 1s}\left(v_1\right)-T_{\rm 1s}\left(v_4\right)$ are applied to the CD4 T lymphocytes (T4), a critical cell type that helps to coordinate the immune response (\Cref{fig: upset}, results for other cell types are presented in \Cref{app: immune cell results}). For $T_{\rm anc}\left(v_1\right)$, the reported p-value corresponds to the global null, whereas the debiased tests $T_{\rm 1s}(v_k)$, $k = 1,\ldots,4$, correspond to the projected nulls $(\mu_X - \mu_Z)^\top v_k = 0$, $k = 1,\ldots,4$, respectively. The anchored test and the PC1, PC4 debiased tests report significant differences at the standard $0.05$ threshold. The latter two results answer our motivating questions in the Introduction (\Cref{fig: motivation}B): the observed distributional difference between case and control samples, in the directions of PC1 and PC4, is indeed statistically significant. 

\begin{figure}[!t]
    \centering
    \includegraphics[width =0.9\linewidth]{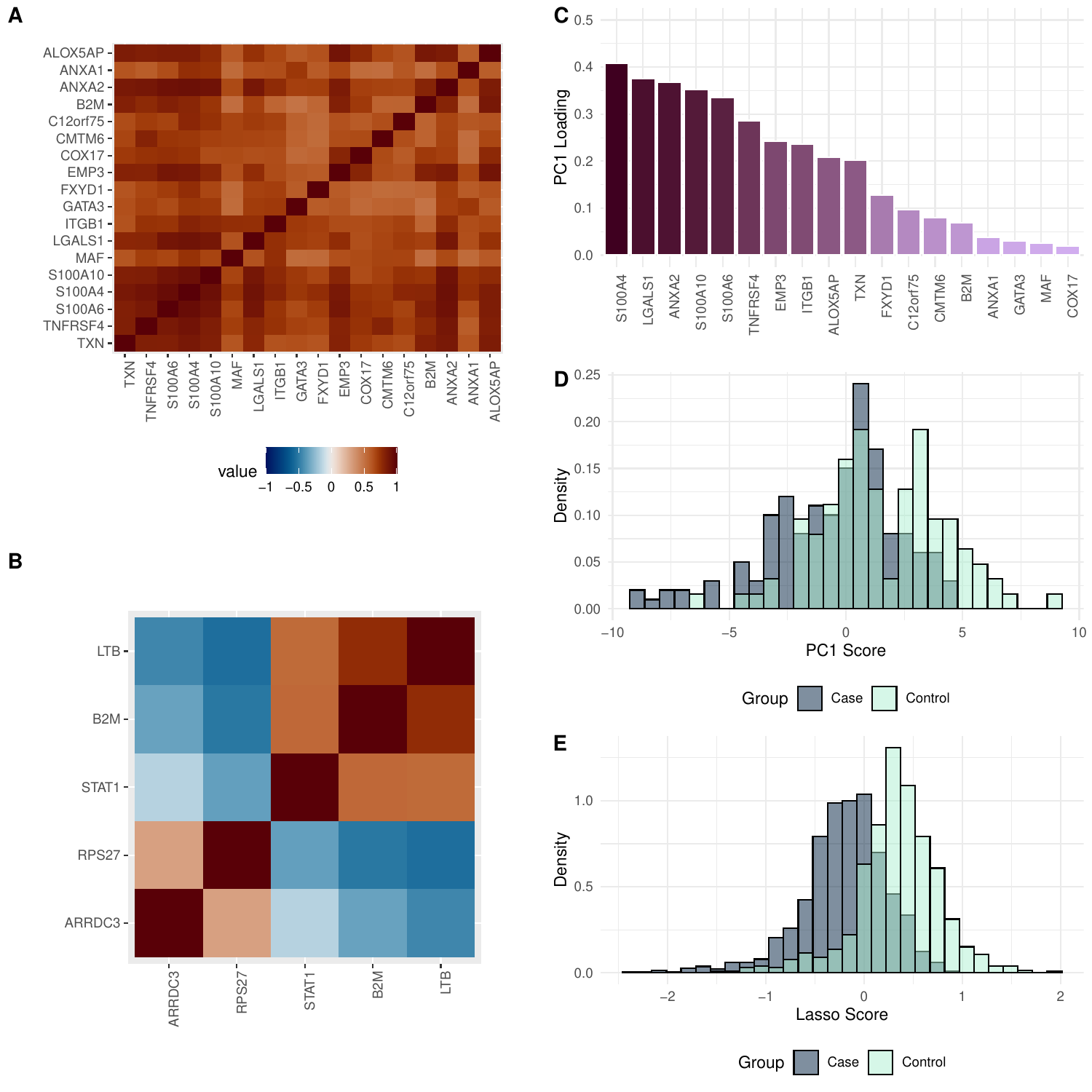}
    \caption{Further examination of the analysis results. \textbf{A}. Sample correlation between genes contained in (estimated) PC1. This is also a zoom-in inspection of the top left block in \Cref{fig: motivation}A. \textbf{B}. Sample correlation between genes contained in the Lasso vector. \textbf{C}. Gene loading values in estimated PC1 $v_1^{(-m)}$ (using one split as an example).  \textbf{D}. The distribution of out-of-sample discriminating score, calculated as follows: For each $X_i,Z_i\in\mathcal{D}^{(m)}$, calculate $X_i^\top v_1^{(-m)}$ or $Z_i^\top v_1^{(-m)}$. Iterate over all $M$ splits and collect all the scores. Present the distribution of all $N_X + N_Z$ score numbers by group. \textbf{E}. A similar score distribution plot for Lasso. The calculation replaces all the $v_1^{(-m)}$ above by $\beta^{(-m)}$.}
    \label{fig: interpretation}
\end{figure}

Further details regarding the systematic signal are displayed in \Cref{fig: upset} and \ref{fig: interpretation}. The PCs define the ``active genes", which are defined as those genes consistently (across different data splits) taking non-zero loadings in $\beta^{(-m)}$ or $v_k^{(-m)}$: specifically, we define a gene to be active if it takes non-zero loadings in more than half of the estimated high-dimensional sparse vectors (``majority voting''). Assessing how much the non-zero genes vary between splits can also offer researchers basic intuition regarding the noise level when estimating $\beta^{(-m)}, v_k^{(-m)}$.   

The number of active genes for each of the PC1-PC4 is approximately $20$ (\Cref{fig: upset}). PC1 ($p=0.047$) includes $18$ genes and $14$ of them are reported to be marginally significant according to a standard univariate negative-binomial regression (p-value threshold $0.05$); however, only one survives adjustment for multiple comparisons. A further inspection of the PC1 active genes and the estimated $v_1^{(-m)}$ is provided in \Cref{fig: interpretation} A, C. These genes are all highly correlated and likely have similar functions in the immune system. PC4 ($p=0.004$) includes 19 genes and $9$ of them are reported marginally significant according to the marginal tests ($0.05$-level) and 8 retain significance after multiple comparison corrections. We provide the gene names and their correlation in \Cref{appfig: inspect PC4}. In panel A, we can observe the $19$ genes are divided into two association blocks: One contains all the mitochondrial genes, which are not protein-coding genes.  These genes are not well studied in the literature and are often removed from such analyses. The other block contains multiple genes having significant functions in the immune system: as many of their names suggest, the "IFI" prefix stands for "InterFeron-Inducible," indicating that these genes are up-regulated in response to interferon signaling, which is an anti-virus mechanism in the human body.

The anchored-Lasso solution is more parsimonious than the PC methods, identifying 5 active genes (\Cref{fig: upset}). Among the 5, one (B2M) overlaps with the estimated PC1 vector, and the rest are not included in the leading PCs. As illustrated by the sample correlation between these genes (\Cref{fig: interpretation}B), they are not highly correlated. Using this small set of signal genes, we can effectively separate the case and control individuals (\Cref{fig: interpretation}E). Compared with the discrimination capacity of PC1 (\Cref{fig: interpretation}D), Lasso's score distribution is visually more bimodal, which is expected as we selected these genes via a label-prediction task. Among the five active genes, four (except for RPS27) are reported in the biomedical literature to encode important proteins in immune response and/or antiviral activity\cite{creus2012multifaceted,tang2021impact, largent2023dysregulated, takeuchi2019genome}. Depending on the specific purpose of the scientific research, a user can decide which test is more relevant to each of their goals.  Regardless, our methods will likely point them to more biologically meaningful signals than a simple global test.

\section{Discussion}
\label{section: discussion}
In this paper, we examined several projection-based procedures for high-dimensional mean comparison. The discussed sparse projection-based tests offer better interpretability and take advantage of the interaction among the signal features. Depending on the specific scientific question and data availability, practitioners may choose the method that best aligns with their analytic goals.

We investigate the one-step statistics $T_{\rm 1s}(v_j)$ corresponding to PC projective directions. There is also potential to develop new procedures targeting other scientifically meaningful alternatives. Although the general semiparametric framework has been extensively studied, evaluating the practicability of specific statistics continues to present rich opportunities for future research.

We envision the plug-in proposal $T_{\rm pi}$ can extend beyond testing two-group means.
Thanks to its conceptual simplicity, this approach may be adapted to other contexts---for instance, assessing whether the effects of two treatments are comparable in high-dimensional outcome settings, such as gene expression studies. Reformulations of \Cref{th: approximate orthogonal} can support flexible, frequentist-valid inference for these broader settings.


\section*{Acknowledgments}
 This project was funded by National Institute of Mental Health grant R01MH123184 and NSF DMS-2015492. We appreciate Jin-Hong Du's assistance in preparing the data presented in \Cref{section: real data}.

\section*{Code Availability}
The code is available as the CRAN package \texttt{HMC}. For source code and documentation, see the GitHub repository:
\url{https://github.com/terrytianyuzhang/HMC}.

\bibliography{main}

\begin{thebibliography}{}

\bibitem[\protect\citeauthoryear{Abid, Zhang, Bagaria, and Zou}{Abid
  et~al.}{2018}]{abid2018exploring}
Abid, A., M.~J. Zhang, V.~K. Bagaria, and J.~Zou (2018).
\newblock Exploring patterns enriched in a dataset with contrastive principal
  component analysis.
\newblock {\em Nature communications\/}~{\em 9\/}(1), 2134.

\bibitem[\protect\citeauthoryear{Ahlmann-Eltze and Huber}{Ahlmann-Eltze and
  Huber}{2023}]{ahlmann2023comparison}
Ahlmann-Eltze, C. and W.~Huber (2023).
\newblock Comparison of transformations for single-cell rna-seq data.
\newblock {\em Nature Methods\/}~{\em 20\/}(5), 665--672.

\bibitem[\protect\citeauthoryear{Bai and Saranadasa}{Bai and
  Saranadasa}{1996}]{bai1996effect}
Bai, Z. and H.~Saranadasa (1996).
\newblock Effect of high dimension: by an example of a two sample problem.
\newblock {\em Statistica Sinica\/}, 311--329.

\bibitem[\protect\citeauthoryear{Bickel, Klaassen, Bickel, Ritov, Klaassen,
  Wellner, and Ritov}{Bickel et~al.}{1993}]{bickel1993efficient}
Bickel, P.~J., C.~A. Klaassen, P.~J. Bickel, Y.~Ritov, J.~Klaassen, J.~A.
  Wellner, and Y.~Ritov (1993).
\newblock {\em Efficient and adaptive estimation for semiparametric models},
  Volume~4.
\newblock Springer.

\bibitem[\protect\citeauthoryear{Cai and Liu}{Cai and
  Liu}{2011}]{cai2011direct}
Cai, T. and W.~Liu (2011).
\newblock A direct estimation approach to sparse linear discriminant analysis.
\newblock {\em Journal of the American statistical association\/}~{\em
  106\/}(496), 1566--1577.

\bibitem[\protect\citeauthoryear{Cai, Ren, and Zhou}{Cai
  et~al.}{2016}]{cai2016review}
Cai, T.~T., Z.~Ren, and H.~H. Zhou (2016).
\newblock {Estimating structured high-dimensional covariance and precision
  matrices: Optimal rates and adaptive estimation}.
\newblock {\em Electronic Journal of Statistics\/}~{\em 10\/}(1), 1 -- 59.

\bibitem[\protect\citeauthoryear{Carvalho, Chang, Lucas, Nevins, Wang, and
  West}{Carvalho et~al.}{2008}]{Carvalho:2008}
Carvalho, C.~M., J.~Chang, J.~E. Lucas, J.~R. Nevins, Q.~Wang, and M.~West
  (2008, Dec).
\newblock High-dimensional sparse factor modeling: Applications in gene
  expression genomics.
\newblock {\em J Am Stat Assoc\/}~{\em 103\/}(484), 1438--1456.

\bibitem[\protect\citeauthoryear{Chen and Qin}{Chen and
  Qin}{2010}]{chen2010two}
Chen, S.~X. and Y.-L. Qin (2010).
\newblock {A two-sample test for high-dimensional data with applications to
  gene-set testing}.
\newblock {\em The Annals of Statistics\/}~{\em 38\/}(2), 808 -- 835.

\bibitem[\protect\citeauthoryear{Chernozhukov, Chetverikov, Demirer, Duflo,
  Hansen, Newey, and Robins}{Chernozhukov et~al.}{2018}]{cher2018doubleml}
Chernozhukov, V., D.~Chetverikov, M.~Demirer, E.~Duflo, C.~Hansen, W.~Newey,
  and J.~Robins (2018, 01).
\newblock {Double/debiased machine learning for treatment and structural
  parameters}.
\newblock {\em The Econometrics Journal\/}~{\em 21\/}(1), C1--C68.

\bibitem[\protect\citeauthoryear{Creus, De~Paepe, Weis, and De~Bleecker}{Creus
  et~al.}{2012}]{creus2012multifaceted}
Creus, K.~K., B.~De~Paepe, J.~Weis, and J.~L. De~Bleecker (2012).
\newblock The multifaceted character of lymphotoxin $\beta$ in inflammatory
  myopathies and muscular dystrophies.
\newblock {\em Neuromuscular Disorders\/}~{\em 22\/}(8), 712--719.

\bibitem[\protect\citeauthoryear{Critchley}{Critchley}{1985}]{critchley1985influence}
Critchley, F. (1985).
\newblock Influence in principal components analysis.
\newblock {\em Biometrika\/}~{\em 72\/}(3), 627--636.

\bibitem[\protect\citeauthoryear{Dai, Shen, and Pan}{Dai
  et~al.}{2022}]{dai2022significance}
Dai, B., X.~Shen, and W.~Pan (2022).
\newblock Significance tests of feature relevance for a black-box learner.
\newblock {\em IEEE Transactions on Neural Networks and Learning Systems\/}.

\bibitem[\protect\citeauthoryear{Dixit, Parnas, Li, Chen, Fulco, Jerby-Arnon,
  Marjanovic, Dionne, Burks, Raychowdhury, et~al.}{Dixit
  et~al.}{2016}]{Dixit2016}
Dixit, A., O.~Parnas, B.~Li, J.~Chen, C.~P. Fulco, L.~Jerby-Arnon, N.~D.
  Marjanovic, D.~Dionne, T.~Burks, R.~Raychowdhury, et~al. (2016).
\newblock Perturb-seq: dissecting molecular circuits with scalable single-cell
  {RNA} profiling of pooled genetic screens.
\newblock {\em Cell\/}~{\em 167\/}(7), 1853--1866.

\bibitem[\protect\citeauthoryear{Donoho}{Donoho}{1995}]{donoho1995noising}
Donoho, D.~L. (1995).
\newblock De-noising by soft-thresholding.
\newblock {\em IEEE transactions on information theory\/}~{\em 41\/}(3),
  613--627.

\bibitem[\protect\citeauthoryear{Fan, Liao, and Liu}{Fan
  et~al.}{2016}]{fan2016overview}
Fan, J., Y.~Liao, and H.~Liu (2016).
\newblock An overview of the estimation of large covariance and precision
  matrices.
\newblock {\em The Econometrics Journal\/}~{\em 19\/}(1), C1--C32.

\bibitem[\protect\citeauthoryear{Fisher and Kennedy}{Fisher and
  Kennedy}{2021}]{fisher2021visually}
Fisher, A. and E.~H. Kennedy (2021).
\newblock Visually communicating and teaching intuition for influence
  functions.
\newblock {\em The American Statistician\/}~{\em 75\/}(2), 162--172.

\bibitem[\protect\citeauthoryear{Glynn and Quinn}{Glynn and
  Quinn}{2010}]{glynn2010introduction}
Glynn, A.~N. and K.~M. Quinn (2010).
\newblock An introduction to the augmented inverse propensity weighted
  estimator.
\newblock {\em Political analysis\/}~{\em 18\/}(1), 36--56.

\bibitem[\protect\citeauthoryear{Hines, Dukes, Diaz-Ordaz, and
  Vansteelandt}{Hines et~al.}{2022}]{hines2022demystifying}
Hines, O., O.~Dukes, K.~Diaz-Ordaz, and S.~Vansteelandt (2022).
\newblock Demystifying statistical learning based on efficient influence
  functions.
\newblock {\em The American Statistician\/}~{\em 76\/}(3), 292--304.

\bibitem[\protect\citeauthoryear{Hotelling}{Hotelling}{1931}]{hotellings}
Hotelling, H. (1931).
\newblock {The Generalization of Student's Ratio}.
\newblock {\em The Annals of Mathematical Statistics\/}~{\em 2\/}(3), 360 --
  378.

\bibitem[\protect\citeauthoryear{Huang, Li, Li, and Yang}{Huang
  et~al.}{2022}]{huang2022overview}
Huang, Y., C.~Li, R.~Li, and S.~Yang (2022).
\newblock An overview of tests on high-dimensional means.
\newblock {\em Journal of multivariate analysis\/}~{\em 188}, 104813.

\bibitem[\protect\citeauthoryear{Jin, Simmons, Guo, Shetty, Ko, Nguyen, Jokhi,
  Robinson, Oyler, Curry, et~al.}{Jin et~al.}{2020}]{jin2020vivo}
Jin, X., S.~K. Simmons, A.~Guo, A.~S. Shetty, M.~Ko, L.~Nguyen, V.~Jokhi,
  E.~Robinson, P.~Oyler, N.~Curry, et~al. (2020).
\newblock In vivo perturb-seq reveals neuronal and glial abnormalities
  associated with autism risk genes.
\newblock {\em Science\/}~{\em 370\/}(6520), eaaz6063.

\bibitem[\protect\citeauthoryear{Johnstone}{Johnstone}{2001}]{johnstone2001distribution}
Johnstone, I.~M. (2001).
\newblock On the distribution of the largest eigenvalue in principal components
  analysis.
\newblock {\em The Annals of statistics\/}~{\em 29\/}(2), 295--327.

\bibitem[\protect\citeauthoryear{Johnstone}{Johnstone}{2019}]{johnstone2019gaussian}
Johnstone, I.~M. (2019).
\newblock Gaussian estimation: Sequence and wavelet models.
\newblock Unpublished Book.

\bibitem[\protect\citeauthoryear{Jones, Townes, Li, and Engelhardt}{Jones
  et~al.}{2022}]{jones2022contrastive}
Jones, A., F.~W. Townes, D.~Li, and B.~E. Engelhardt (2022).
\newblock Contrastive latent variable modeling with application to case-control
  sequencing experiments.
\newblock {\em The Annals of Applied Statistics\/}~{\em 16\/}(3), 1268--1291.

\bibitem[\protect\citeauthoryear{Kennedy}{Kennedy}{2022}]{kennedy2022semiparametric}
Kennedy, E.~H. (2022).
\newblock Semiparametric doubly robust targeted double machine learning: a
  review.
\newblock {\em arXiv preprint arXiv:2203.06469\/}.

\bibitem[\protect\citeauthoryear{Knowles and Ghahramani}{Knowles and
  Ghahramani}{2011}]{Knowles:2011}
Knowles, D. and Z.~Ghahramani (2011).
\newblock {Nonparametric Bayesian sparse factor models with application to gene
  expression modeling}.
\newblock {\em The Annals of Applied Statistics\/}~{\em 5\/}(2B), 1534 -- 1552.

\bibitem[\protect\citeauthoryear{Kosorok}{Kosorok}{2008}]{kosorok2008introduction}
Kosorok, M.~R. (2008).
\newblock {\em Introduction to empirical processes and semiparametric
  inference}, Volume~61.
\newblock Springer.

\bibitem[\protect\citeauthoryear{Lam}{Lam}{2020}]{lam2020high}
Lam, C. (2020).
\newblock High-dimensional covariance matrix estimation.
\newblock {\em Wiley Interdisciplinary reviews: computational
  statistics\/}~{\em 12\/}(2), e1485.

\bibitem[\protect\citeauthoryear{Langfelder and Horvath}{Langfelder and
  Horvath}{2007}]{langfelder2007eigengene}
Langfelder, P. and S.~Horvath (2007).
\newblock Eigengene networks for studying the relationships between
  co-expression modules.
\newblock {\em BMC systems biology\/}~{\em 1}, 1--17.

\bibitem[\protect\citeauthoryear{Langfelder and Horvath}{Langfelder and
  Horvath}{2008}]{langfelder2008wgcna}
Langfelder, P. and S.~Horvath (2008).
\newblock Wgcna: an r package for weighted correlation network analysis.
\newblock {\em BMC bioinformatics\/}~{\em 9\/}(1), 1--13.

\bibitem[\protect\citeauthoryear{Largent, Lambert, Chiang, Shumlak, Liggitt,
  Oukka, Torgerson, Buckner, Allenspach, Rawlings, et~al.}{Largent
  et~al.}{2023}]{largent2023dysregulated}
Largent, A.~D., K.~Lambert, K.~Chiang, N.~Shumlak, D.~Liggitt, M.~Oukka, T.~R.
  Torgerson, J.~H. Buckner, E.~J. Allenspach, D.~J. Rawlings, et~al. (2023).
\newblock Dysregulated ifn-$\gamma$ signals promote autoimmunity in stat1
  gain-of-function syndrome.
\newblock {\em Science Translational Medicine\/}~{\em 15\/}(703), eade7028.

\bibitem[\protect\citeauthoryear{Liu, Yu, and Li}{Liu
  et~al.}{2022}]{liu2022multiple}
Liu, W., X.~Yu, and R.~Li (2022).
\newblock Multiple-splitting projection test for high-dimensional mean vectors.
\newblock {\em The Journal of Machine Learning Research\/}~{\em 23\/}(1),
  3091--3117.

\bibitem[\protect\citeauthoryear{Lucas, Kung, and Chi}{Lucas
  et~al.}{2010}]{Lucas:2010}
Lucas, J.~E., H.-N. Kung, and J.-T.~A. Chi (2010, Sep).
\newblock Latent factor analysis to discover pathway-associated putative
  segmental aneuploidies in human cancers.
\newblock {\em PLoS Comput Biol\/}~{\em 6\/}(9), e1000920.

\bibitem[\protect\citeauthoryear{Lundborg, Kim, Shah, and Samworth}{Lundborg
  et~al.}{2022}]{lundborg2022projected}
Lundborg, A.~R., I.~Kim, R.~D. Shah, and R.~J. Samworth (2022).
\newblock The projected covariance measure for assumption-lean variable
  significance testing.
\newblock {\em arXiv preprint arXiv:2211.02039\/}.

\bibitem[\protect\citeauthoryear{Magnus}{Magnus}{1985}]{magnus1985differentiating}
Magnus, J.~R. (1985).
\newblock On differentiating eigenvalues and eigenvectors.
\newblock {\em Econometric theory\/}~{\em 1\/}(2), 179--191.

\bibitem[\protect\citeauthoryear{Perez, Gordon, Subramaniam, Kim, Hartoularos,
  Targ, Sun, Ogorodnikov, Bueno, Lu, et~al.}{Perez
  et~al.}{2022}]{perez2022single}
Perez, R.~K., M.~G. Gordon, M.~Subramaniam, M.~C. Kim, G.~C. Hartoularos,
  S.~Targ, Y.~Sun, A.~Ogorodnikov, R.~Bueno, A.~Lu, et~al. (2022).
\newblock Single-cell rna-seq reveals cell type--specific molecular and genetic
  associations to lupus.
\newblock {\em Science\/}~{\em 376\/}(6589), eabf1970.

\bibitem[\protect\citeauthoryear{Rakshit, Wang, Cai, and Guo}{Rakshit
  et~al.}{2023}]{rakshit2023sihr}
Rakshit, P., Z.~Wang, T.~T. Cai, and Z.~Guo (2023).
\newblock Sihr: Statistical inference in high-dimensional linear and logistic
  regression models.

\bibitem[\protect\citeauthoryear{Replogle, Saunders, Pogson, Hussmann, Lenail,
  Guna, Mascibroda, Wagner, Adelman, Lithwick-Yanai, Iremadze, Oberstrass,
  Lipson, Bonnar, Jost, Norman, and Weissman}{Replogle
  et~al.}{2022}]{Replogle2022}
Replogle, J.~M., R.~A. Saunders, A.~N. Pogson, J.~A. Hussmann, A.~Lenail,
  A.~Guna, L.~Mascibroda, E.~J. Wagner, K.~Adelman, G.~Lithwick-Yanai,
  N.~Iremadze, F.~Oberstrass, D.~Lipson, J.~L. Bonnar, M.~Jost, T.~M. Norman,
  and J.~S. Weissman (2022).
\newblock Mapping information-rich genotype-phenotype landscapes with
  genome-scale perturb-seq.
\newblock {\em Cell\/}~{\em 185\/}(14), 2559--2575.e28.

\bibitem[\protect\citeauthoryear{Schraivogel, Gschwind, Milbank, Leonce, Jakob,
  Mathur, Korbel, Merten, Velten, and Steinmetz}{Schraivogel
  et~al.}{2020}]{Schraivogel2020}
Schraivogel, D., A.~R. Gschwind, J.~H. Milbank, D.~R. Leonce, P.~Jakob,
  L.~Mathur, J.~O. Korbel, C.~A. Merten, L.~Velten, and L.~M. Steinmetz (2020).
\newblock Targeted perturb-seq enables genome-scale genetic screens in single
  cells.
\newblock {\em Nature Methods\/}~{\em 17\/}(6), 629--635.

\bibitem[\protect\citeauthoryear{Shao, Wang, Deng, and Wang}{Shao
  et~al.}{2011}]{shao2011LDA}
Shao, J., Y.~Wang, X.~Deng, and S.~Wang (2011).
\newblock {Sparse linear discriminant analysis by thresholding for high
  dimensional data}.
\newblock {\em The Annals of Statistics\/}~{\em 39\/}(2), 1241 -- 1265.

\bibitem[\protect\citeauthoryear{Stein}{Stein}{1956}]{stein1956inadmissibility}
Stein, C. (1956).
\newblock Inadmissibility of the usual estimator for the mean of a multivariate
  normal distribution.
\newblock In {\em Proceedings of the Third Berkeley Symposium on Mathematical
  Statistics and Probability, Volume 1: Contributions to the Theory of
  Statistics}, Volume~3, pp.\  197--207. University of California Press.

\bibitem[\protect\citeauthoryear{Stein-O'Brien, Arora, Culhane, Favorov,
  Garmire, Greene, Goff, Li, Ngom, Ochs, Xu, and Fertig}{Stein-O'Brien
  et~al.}{2018}]{Stein-OBrien:2018}
Stein-O'Brien, G.~L., R.~Arora, A.~C. Culhane, A.~V. Favorov, L.~X. Garmire,
  C.~S. Greene, L.~A. Goff, Y.~Li, A.~Ngom, M.~F. Ochs, Y.~Xu, and E.~J. Fertig
  (2018, Oct).
\newblock Enter the matrix: Factorization uncovers knowledge from omics.
\newblock {\em Trends Genet\/}~{\em 34\/}(10), 790--805.

\bibitem[\protect\citeauthoryear{Stewart}{Stewart}{1977}]{stewart1977perturbation}
Stewart, G.~W. (1977).
\newblock On the perturbation of pseudo-inverses, projections and linear least
  squares problems.
\newblock {\em SIAM review\/}~{\em 19\/}(4), 634--662.

\bibitem[\protect\citeauthoryear{Stuart, Segal, Koller, and Kim}{Stuart
  et~al.}{2003}]{stuart2003gene}
Stuart, J.~M., E.~Segal, D.~Koller, and S.~K. Kim (2003).
\newblock A gene-coexpression network for global discovery of conserved genetic
  modules.
\newblock {\em science\/}~{\em 302\/}(5643), 249--255.

\bibitem[\protect\citeauthoryear{Su, Xu, Shan, Cai, Zhao, and Zhang}{Su
  et~al.}{2023}]{su2023cell}
Su, C., Z.~Xu, X.~Shan, B.~Cai, H.~Zhao, and J.~Zhang (2023).
\newblock Cell-type-specific co-expression inference from single cell
  rna-sequencing data.
\newblock {\em Nature Communications\/}~{\em 14\/}(1), 4846.

\bibitem[\protect\citeauthoryear{Takeuchi, Kukimoto, Li, Li, Li, Hu, Takahashi,
  Inoue, Yokoi, Chen, et~al.}{Takeuchi et~al.}{2019}]{takeuchi2019genome}
Takeuchi, F., I.~Kukimoto, Z.~Li, S.~Li, N.~Li, Z.~Hu, A.~Takahashi, S.~Inoue,
  S.~Yokoi, J.~Chen, et~al. (2019).
\newblock Genome-wide association study of cervical cancer suggests a role for
  arrdc3 gene in human papillomavirus infection.
\newblock {\em Human molecular genetics\/}~{\em 28\/}(2), 341--348.

\bibitem[\protect\citeauthoryear{Tang, Barbacioru, Wang, Nordman, Lee, Xu,
  Wang, Bodeau, Tuch, Siddiqui, et~al.}{Tang et~al.}{2009}]{tang2009mrna}
Tang, F., C.~Barbacioru, Y.~Wang, E.~Nordman, C.~Lee, N.~Xu, X.~Wang,
  J.~Bodeau, B.~B. Tuch, A.~Siddiqui, et~al. (2009).
\newblock mrna-seq whole-transcriptome analysis of a single cell.
\newblock {\em Nature methods\/}~{\em 6\/}(5), 377--382.

\bibitem[\protect\citeauthoryear{Tang, Zhao, Zhang, Wei, Tian, Li, Yao, Wang,
  and Li}{Tang et~al.}{2021}]{tang2021impact}
Tang, F., Y.-H. Zhao, Q.~Zhang, W.~Wei, S.-F. Tian, C.~Li, J.~Yao, Z.-F. Wang,
  and Z.-Q. Li (2021).
\newblock Impact of beta-2 microglobulin expression on the survival of glioma
  patients via modulating the tumor immune microenvironment.
\newblock {\em CNS Neuroscience \& Therapeutics\/}~{\em 27\/}(8), 951--962.

\bibitem[\protect\citeauthoryear{Tibshirani}{Tibshirani}{1996}]{tibshirani1996regression}
Tibshirani, R. (1996).
\newblock Regression shrinkage and selection via the lasso.
\newblock {\em Journal of the Royal Statistical Society Series B: Statistical
  Methodology\/}~{\em 58\/}(1), 267--288.

\bibitem[\protect\citeauthoryear{Tony~Cai, Liu, and Xia}{Tony~Cai
  et~al.}{2014}]{tony2014two}
Tony~Cai, T., W.~Liu, and Y.~Xia (2014).
\newblock Two-sample test of high dimensional means under dependence.
\newblock {\em Journal of the Royal Statistical Society Series B: Statistical
  Methodology\/}~{\em 76\/}(2), 349--372.

\bibitem[\protect\citeauthoryear{Tsiatis}{Tsiatis}{2006}]{tsiatis2006semiparametric}
Tsiatis, A.~A. (2006).
\newblock Semiparametric theory and missing data.

\bibitem[\protect\citeauthoryear{van~de Geer}{van~de Geer}{2008}]{van2008high}
van~de Geer, S.~A. (2008).
\newblock High-dimensional generalized linear models and the lasso.
\newblock {\em The Annals of Statistics\/}~{\em 36\/}(2), 614--645.

\bibitem[\protect\citeauthoryear{Wainwright}{Wainwright}{2019}]{wainwright2019high}
Wainwright, M.~J. (2019).
\newblock {\em High-dimensional statistics: A non-asymptotic viewpoint},
  Volume~48.
\newblock Cambridge university press.

\bibitem[\protect\citeauthoryear{Wedin}{Wedin}{1973}]{wedin1973perturbation}
Wedin, P.-{\AA}. (1973).
\newblock Perturbation theory for pseudo-inverses.
\newblock {\em BIT Numerical Mathematics\/}~{\em 13}, 217--232.

\bibitem[\protect\citeauthoryear{Williamson, Gilbert, Simon, and
  Carone}{Williamson et~al.}{2023}]{williamson2023general}
Williamson, B.~D., P.~B. Gilbert, N.~R. Simon, and M.~Carone (2023).
\newblock A general framework for inference on algorithm-agnostic variable
  importance.
\newblock {\em Journal of the American Statistical Association\/}~{\em
  118\/}(543), 1645--1658.

\bibitem[\protect\citeauthoryear{Witten, Tibshirani, Gross, and
  Narasimhan}{Witten et~al.}{2023}]{PMA}
Witten, D., R.~Tibshirani, S.~Gross, and B.~Narasimhan (2023).
\newblock {\em PMA: Penalized Multivariate Analysis}.
\newblock R package version 1.2-2.

\bibitem[\protect\citeauthoryear{Wolf, Angerer, and Theis}{Wolf
  et~al.}{2018}]{wolf2018scanpy}
Wolf, F.~A., P.~Angerer, and F.~J. Theis (2018).
\newblock Scanpy: large-scale single-cell gene expression data analysis.
\newblock {\em Genome biology\/}~{\em 19}, 1--5.

\bibitem[\protect\citeauthoryear{Yao, Binan, Bezney, Simonton, Freedman,
  Frangieh, Dey, Geiger-Schuller, Eraslan, Gusev, et~al.}{Yao
  et~al.}{2024}]{yao2024scalable}
Yao, D., L.~Binan, J.~Bezney, B.~Simonton, J.~Freedman, C.~J. Frangieh, K.~Dey,
  K.~Geiger-Schuller, B.~Eraslan, A.~Gusev, et~al. (2024).
\newblock Scalable genetic screening for regulatory circuits using compressed
  perturb-seq.
\newblock {\em Nature biotechnology\/}~{\em 42\/}(8), 1282--1295.

\bibitem[\protect\citeauthoryear{Yu, Wang, and Samworth}{Yu
  et~al.}{2015}]{yu2015useful}
Yu, Y., T.~Wang, and R.~J. Samworth (2015).
\newblock A useful variant of the davis--kahan theorem for statisticians.
\newblock {\em Biometrika\/}~{\em 102\/}(2), 315--323.

\bibitem[\protect\citeauthoryear{Zhou, Luo, Liang, Chen, and He}{Zhou
  et~al.}{2023}]{zhou2023new}
Zhou, Y., K.~Luo, L.~Liang, M.~Chen, and X.~He (2023).
\newblock A new bayesian factor analysis method improves detection of genes and
  biological processes affected by perturbations in single-cell crispr
  screening.
\newblock {\em Nature Methods\/}.

\bibitem[\protect\citeauthoryear{Zou, Hsu, Parkes, and Adams}{Zou
  et~al.}{2013}]{zou2013contrastive}
Zou, J.~Y., D.~J. Hsu, D.~C. Parkes, and R.~P. Adams (2013).
\newblock Contrastive learning using spectral methods.
\newblock {\em Advances in Neural Information Processing Systems\/}~{\em 26}.

\end{thebibliography}
\bibliographystyle{chicago}

\newpage
\appendix
\renewcommand\thefigure{\thesection.\arabic{figure}}    
\section{Simulation settings in \Cref{fig:cross-fitting under null}}
\label{app: for the figure}
The distribution of $T_{\rm pi}$ under the global and projected nulls is presented in \Cref{fig:cross-fitting under null}. Here we present the details of the simulation settings.

The two-group samples $\{X_i\}, \{Z_i\}$ are IID multivariate normal with equal covariance matrix. We used a two-split crossing fitting ($M=2$). Number of samples in each split: $n_X = 250$, $n_Z = 125$. Dimension of $X,Z$: $p = 100$. The mean of $X$ is
\begin{equation*}
    \mu_X = (\underbrace{2.5,...,2.5}_{5}, \underbrace{-2.5,...,-2.5}_{5}, \underbrace{0,...,0}_{90}).
\end{equation*}
And that of $Z$ is 
\begin{equation*}
    \mu_Z = (\underbrace{-2.5,...,-2.5}_{5}, \underbrace{2.5,...,2.5}_{5}, \underbrace{0,...,0}_{90}).
\end{equation*}
The covariance matrices of $X,Z$ are:
\begin{equation*}
    \Sigma_X = \Sigma_Z = 3v_1v_1^\top+I_p,
\end{equation*}
where $I_p$ is $p$-dimensional identity matrix. The top PC vector $v$ is:
\begin{equation*}
    v_1 = (\underbrace{0.316, ..., 0.316}_{10},\underbrace{0,...,0}_{90}).
\end{equation*}
Note that we normalized $v_1$ such that $\|v_1\| = 1$.

For this, we implement standard PCA to estimate $v_1$. We are aware that standard PCA is not a consistent estimator of $v_1$ in the high-dimensional setting but still stick to this choice because standard PCA is still routinely applied in high-dimensional biomedical research. Statistically, it is not the best practice but in this specific simulation, the estimation quality is satisfactory. Results do not significantly change after switching to sPCA.

\section{Explicit Formulas for the Debiased Test}\label{app: explicit formula}

We omitted to present the explicit formula of several quantities for constructing $T_{\rm 1s}$ in the main text to save some space. We present them in this section.

For simplicity of notation, we use $P_n^{(m)}(\cdot)$ to denote ``taking empirical averaging with $\mathcal{D}^{(m)}$''. For example,
\begin{equation*}
    P_n^{(m)} (X^\top v^{(-m)}) := n_X^{-1}\sum_{X_i\in\mathcal{D}^{(m)}} X_i^\top v^{(-m)} = \mu_X^{(m)\top}v^{(-m)}.
\end{equation*}

The population-level influence functions, $\phi_X(X)$ and $\phi_Z(Z)$, of the eigenvector $v_1$ functional  are:
\begin{equation}\label{eq: define_true_influence}
    \begin{aligned}
\phi_X(X) & =s^{\top}\left[\left(X-\mu_X\right)\left(X-\mu_X \right)^{\top}-\Sigma \right] v_1 , \\
\phi_Z (Z) & =s^{ \top}\left[\left(Z-\mu_Z \right)\left(Z-\mu_Z \right)^{\top}-\Sigma \right] v_1 , \\
s  & =\left(\lambda_1 I_p-\Sigma \right)^{+}\left(\mu_X -\mu_Z \right) .
\end{aligned}
\end{equation}

The variance estimator $\hat{\sigma}_{\rm 1s}^2$ in $T_{\rm 1s}$ is 
\begin{equation}\label{eq:debiased test standard error}
        \sum_{m=1}^M\left\{n_X^{-1} P_n^{(m)}\left[V_X^{(-m)}(X)-P_n^{(m)} V_X^{(-m)}(X)\right]^2+n_Z^{-1} P_n^{(m)}\left[V_Z^{(-m)}(Z)-P_n^{(m)} V^{(-m)}_Z(Z)\right]^2\right\}   
\end{equation}
where
\begin{equation*}
    \begin{aligned}
        V_X^{(-m)}(X)& = \left(X-\mu_X^{(-m)}\right)^{\top} v_1^{(-m)}+w \phi_X^{(-m)}(X)\\
        V_Z^{(-m)}(Z)& =\left(Z-\mu_Z^{(-m)}\right)^{\top} v_1^{(-m)}-(1-w) \phi_Z^{(-m)}(Z)
    \end{aligned}
\end{equation*}

\section{Proof of Theorem \ref{th: one step}}
\label{app: debiased stat}

In this section, we present the proof of \Cref{th: one step}. We need to decompose the debiased test statistics into a sum of the central limit theorem terms, the empirical process ``cross terms'' and the (Taylor expansion) ``remainder terms''. The latter two are of higher order and do not impact the distribution of the quantity of interest asymptotically (shown in \Cref{lemma: cross term} and \ref{lemma: remainder term}). For notational simplicity, we will drop the subscript of $v_1$ and $v_1^{(-m)}$. We will use $P_n^{(m)}(\cdot)$ to denote ``taking empirical average with respect to data $\mathcal{D}^{(m)}$''. We also use $P^{(m)}(\cdot)$ to denote taking expectation with respect to the underlying distribution $(P_X,P_Z)$, conditioned on $\mathcal{D}^{(-m)}$. For example,
\begin{align}
    P_n^{(m)} (X^\top v^{(-m)}) &:= n_X^{-1}\sum_{X_i\in\mathcal{D}^{(m)}} X_i^\top v^{(-m)} = \mu_X^{(m)\top}v^{(-m)}\nonumber\\
    P^{(m)}(X^\top v^{(-m)}) & := E[X^\top v^{(-m)}\mid \mathcal{D}^{(-m)}] = \mu_X^\top v^{(-m)}\nonumber.
\end{align}
Define
\begin{align*}
    \phi(X,Z) & := \frac{n_X}{n_X + n_Z} \phi_X(X) + \frac{n_Z}{n_X + n_Z} \phi_Z(Z)\\
    \phi^{(-m)}(X,Z) & := \frac{n_X}{n_X + n_Z} \phi_X^{(-m)}(X) + \frac{n_Z}{n_X + n_Z} \phi_Z^{(-m)}(Z)\\
    w &= n_X/(n_X + n_Z).
\end{align*}
To clarify, the notation $P_n^{(m)}\phi^{(-m)}(X,Z)$ means
\begin{align*}
    P_n^{(m)}\phi^{(-m)}(X,Z) 
    &= \frac{n_X}{n_X + n_Z} P_n^{(m)}\phi_X^{(-m)}(X) + \frac{n_Z}{n_X + n_Z} P_n^{(m)}\phi_Z^{(-m)}(Z)\\
    & = (n_X+n_Z)^{-1}\left\{\sum_{X_i \in \mathcal{D}^{(m)}} \phi_X^{(-m)}\left(X_i\right)+\sum_{Z_i \in \mathcal{D}^{(m)}} \phi_Z^{(-m)}\left(Z_i\right)\right\}.
\end{align*}
\begin{proof}(Proof of \Cref{th: one step})
For each one of the splits, we will decompose its debiased estimate $\hat \theta_{\rm 1s}^{(m)}$ of $\theta$ into the aforementioned three terms and analyze them separately. The following step is merely algebra, we don't need any assumptions on $\theta$.
    \begin{equation}
    \label{eq: basic decomposition for debiased}
    \begin{aligned}
      & P_n^{(m)}\left((X-Z)^{\top} v^{(-m)}+\phi^{(-m)}(X, Z)\right)  \\
      &  = (P_n^{(m)} - P^{(m)})((X-Z)^{\top} v+\phi (X, Z)) +\\
      &  (P_n^{(m)} - P^{(m)})\left((X-Z)^{\top} v^{(-m)}+\phi^{(-m)}(X, Z) - (X-Z)^{\top} v -\phi (X, Z)\right) + \\
      &  P^{(m)}\left((X-Z)^{\top} v^{(-m)}+\phi^{(-m)}(X, Z)\right).
    \end{aligned}
\end{equation}
The first term in \eqref{eq: basic decomposition for debiased} is the main term that converges to a normal distribution, we will analyze its behavior soon. The vanishing latter two terms are handled in \Cref{lemma: cross term} and \ref{lemma: remainder term}. 

The summation of the estimate over $M$ splits can be written as:
\begin{equation}
\label{eq: debias for lindeberg}
\begin{aligned}
\sum_{m=1}^{M} \hat \theta_{\rm 1s}^{(m)}
      & = \sum_{m=1}^M
\left(P_n^{(m)}-P^{(m)}\right)\left((X-Z)^{\top} v+\phi(X, Z)\right) + \text{higher order terms}\\
& = \left\{\sum_{i=1}^{N_X}
n_X^{-1}\left(X_i-\mu_X\right)^{\top} v + 
(n_X+n_Z)^{-1}\phi_X(X_i) \right\}-\\
& \quad \left\{\sum_{i=1}^{N_Z}
n_Z^{-1}\left(Z_i-\mu_Z\right)^{\top} v - 
(n_X+n_Z)^{-1}\phi_Z(Z_i)\right\}+ o_P\left(n^{-1 / 2}\right).
\end{aligned}
\end{equation}
Note the influence function is mean-zero at the true distribution: $P^{(m)}(\phi(X,Z)) = E[\phi(X,Z)] = 0$.

We need to normalize the summation in \eqref{eq: debias for lindeberg} to apply Lindeberg's central limit theorem. The variance of the main terms in \eqref{eq: debias for lindeberg} is
\begin{equation}
\label{eq: debias lindeberg variance}
\begin{aligned}
    \sigma^2_{\rm 1s} & = M n_X^{-1} \operatorname{Var}\left\{\left(X-\mu_X\right)^{\top} v+w\phi_X(X)\right\}+ \\
    & M n_Z^{-1} \operatorname{Var}\left\{\left(Z-\mu_Z\right)^{\top} v - (1-w) \phi_Z(Z)\right\}.
\end{aligned}
\end{equation}
We also note that Lindeberg's condition is satisfied because the summands have finite second moments,

Our proposal \eqref{eq:debiased test standard error} used a consistent estimator $\hat \sigma_{\rm 1s}^2$ of $\sigma_{\rm 1s}^2$. The testing statistics
\begin{align*}
   T_{\rm 1s} 
        & = \hat \sigma_{\rm 1s}^{-1}\sum_{m=1}^M P_n^{(m)}\left((X-Z)^{\top} v^{(-m)}+\phi^{(-m)}(X, Z)\right)\\
        & = (\sigma_{\rm 1s}/\hat \sigma_{\rm 1s})\sigma_{\rm 1s}^{-1}\sum_{m=1}^M\left(P_n^{(m)}-P\right)\left((X-Z)^{\top} v +\phi(X, Z)\right)\\
        & \quad +(\sigma_{\rm 1s}/\hat \sigma_{\rm 1s})o_P\left(\sigma_{\rm 1s}^{-1}n^{-1 / 2}\right)
        \rightarrow \mathcal{N}(0,1). 
\end{align*}

Note that $\sigma_{\rm 1s}^{-1}$ diverges no faster than $n^{1/2}$. We verify that $\hat \sigma_{\rm 1s}$ is a consistent estimator of $\sigma_{\rm 1s}$ in \Cref{lemma: variance consistency}. 
\end{proof}

\begin{lemma}\label{lemma: variance consistency} Under the same assumptions as in \Cref{th: one step}, we have
\begin{equation*}
\sigma_{\rm 1s} / \hat{\sigma}_{\rm 1s} \xrightarrow{P} 1.
\end{equation*}
\end{lemma}
\begin{proof}
The definitions of $\sigma_{\rm 1s}$ and $\hat \sigma_{\rm 1s}$ are given in \eqref{eq: debias lindeberg variance} and \eqref{eq:debiased test standard error} respectively. It is equivalent to show that $(\hat{\sigma}_{\rm 1s}^2 - \sigma^2_{\rm 1s}) / \sigma^2_{\rm 1s}$ converges to $0$. Note that $\sigma_{\rm 1s}^2 \geq Cn^{-1}$ for some constant $C$ under the assumption that $\operatorname{Var}\left\{\left(X-\mu_X\right)^{\top} v+w \phi_X(X)\right\}$ and $\operatorname{Var}\left\{\left(Z-\mu_Z\right)^{\top} v-(1-w) \phi_Z(Z)\right\}$ are both bounded away from zero. So it is sufficient to show
\begin{equation}\label{eq: variance goal for X}
 \operatorname{Var}\left\{\left(X-\mu_X\right)^{\top} v+w \phi_X(X)\right\} - P_n^{(m)}\left[V_X^{(-m)}(X)-P_n^{(m)} V_X^{(-m)}(X)\right]^2 \rightarrow 0,
\end{equation}
plus 
\begin{equation*}
 \operatorname{Var}\left\{\left(Z-\mu_Z\right)^{\top} v-(1-w) \phi_Z(Z)\right\} - \left.P_n^{(m)}\left[V_Z^{(-m)}(Z)-P_n^{(m)} V_Z^{(-m)}(Z)\right]^2\right\} \rightarrow 0.
\end{equation*}
We focus on \eqref{eq: variance goal for X} and a similar argument holds for $Z$. Denote 
\begin{equation*}
V_X(X)=\left(X-\mu_X\right)^{\top} v+w \phi_X(X)
\end{equation*}
We can decompose \eqref{eq: variance goal for X} as a summation of 
\begin{equation*}
\mathcal{A}=\left(E V_X(X)^2-\left(E V_X(X)\right)^2\right)-\left(P_n^{(m)} V_X(X)^2-\left(P_n^{(m)} V_X(X)\right)^2\right)
\end{equation*}
and 
\begin{equation*}
\mathcal{B}=\left(P_n^{(m)} V_X(X)^2-\left(P_n^{(m)} V_X(X)\right)^2\right)-\left(P_n^{(m)} V_X^{(-m)}(X)^2-\left(P_n^{(m)} V_X^{(-m)}(X)\right)^2\right) .
\end{equation*}
To bound $\mathcal{A}$, we split it into
\begin{equation}\label{eq: split A}
\begin{aligned}
|\mathcal{A}| & \leq\left|P_n^{(m)} V_X^2(X)-E V_X^2(X)\right|+\left|\left(P_n^{(m)} V_X(X)\right)^2-\left(E V_X(X)\right)^2\right| \\
& =\left|P_n^{(m)} V^2_X(X)-E V^2_X(X)\right|+\left|\left(P_n^{(m)} V_X(X)+E V_X(X)\right)\left(P_n^{(m)} V_X(X)-E V_X(X)\right)\right|
\end{aligned}
\end{equation}
We assumed uniform integrability for $\{\left((X_{n1}-\mu_{Xn}\right)^{\top} v_{1n})^2\}$ and $\{(\phi_{Xn}(X_{n1}))^2\}$, this implies finite $EV_{Xn}(X_{n1})^2$ (recall that the analysis is done in a triangular array setting where for different $n$, the samples are generated from a different distribution). Applying Chebyshev's inequality, we know the second term in \eqref{eq: split A} is $|O_P(1) \cdot o_P(1)|$.

It is possible to show $\{V^2_{Xn}(X_{n1})\}$ is also uniformly integrable, noting that $V_X^2(X) \leq 2(\left(X-\mu_X\right)^{\top} v)^2 + 2(\phi_X^2(X))$. To rigorously show that the first term in \Cref{eq: split A} is also $o_P(1)$, we need to apply the triangular array law of large numbers stated in \Cref{lemma: triangular LLN}.

To bound $\mathcal{B}$, we first study
\begin{equation*}
\mathcal{C}=P_n^{(m)} V_X(X)^2-P_n^{(m)} V_X^{(-m)}(X)^2 .
\end{equation*}
And leave 
\begin{equation*}
\mathcal{E}=\left(P_n^{(m)} V_X^{(-m)}(X)\right)^2-\left(P_n^{(m)} V_X(X)\right)^2
\end{equation*}
later.
Apply Cauchy-Schwarz,
\begin{equation*}
    \mathcal{C}^2 \leq P_n^{(m)}\{V_X(X) - V_X^{(-m)}(X)\}^2 \cdot P_n^{(m)} \{V_X(X) + V_X^{(-m)}(X)\}^2 =: \mathcal{F}\cdot \mathcal{G}.
\end{equation*}
We further bound $\mathcal{F}$ by
\begin{equation*}
    \mathcal{F} \leq 2P_n^{(m)}\{(X - \mu_X)^\top v - (X - \mu_X^{(-m)})^\top v^{(-m)}\}^2 + 2w^2P_n^{(m)}\{\phi_X(X) - \phi_X^{(-m)}(X)\}^2.
\end{equation*}
By our assumption $E\left[\left(\phi_X^{(-m)}(X)-\phi_X(X)\right)^2\right]\rightarrow 0$, $P_n^{(m)}\left\{\phi_X(X)-\phi_X^{(-m)}(X)\right\}^2$ is $o_P(1)$. We bound the $\left\{\left(X-\mu_X\right)^{\top} v-\left(X-\mu_X^{(-m)}\right)^{\top} v^{(-m)}\right\}^2$ term in $\mathcal{F}$ by $2$ times
\begin{equation*}
    \left\{\left(X-\mu_X\right)^{\top}\left(v-v^{(-m)}\right)\right\}^2+\left\{\left(\mu_X-\mu_X^{(-m)}\right)^{\top} v^{(-m)}\right\}^2.
\end{equation*}
Each of the two items above has vanishing expectation as $n\rightarrow\infty$ given bounded $\lambda_1(\Sigma_X)$ and $E\left\|v-v^{(-m)}\right\|^2 \rightarrow 0$. Also note that $\|v^{(-m)}\| =1$. So we conclude $\mathcal{F} = o_P(1)$. Applying an almost identical argument, we can show that $\mathcal{G}$ is $O_P(1)$. So we conclude $\mathcal{C}$ is $o_P(1)$ as well.

The bound on $\mathcal{E}$ is implied from that of $\mathcal{C}$:
\begin{equation*}
\begin{aligned}
\mathcal{E}&=P_n^{(m)}\left(V_X(X)-V_X^{(-m)}(X)\right) \cdot P_n^{(m)}\left(V_X(X)+V_X^{(-m)}(X)\right) \\
&\leq \sqrt{P_n^{(m)}\left(V_X(X)-V_X^{(-m)}(X)\right)^2} \cdot \sqrt{P_n^{(m)}\left(V_X(X)+V_X^{(-m)}(X)\right)^2} \\
&=\sqrt{\mathcal{F} \cdot \mathcal{G}}.
\end{aligned}
\end{equation*}
\end{proof}

\begin{lemma}\label{lemma: triangular LLN}Let $P_n, n\in\mathbb{Z}^+$ be a collection of distributions. For each $n$, let $\left\{X_{n k}\right\}_{k=1}^n$ be IID samples from distribution $P_n$. Assume $\{P_n\}$ is uniformly integrable:
\begin{equation*}
\lim _{t \rightarrow \infty} \sup _n E\left[\left|X_{n 1}\right| \cdot \mathbf{1}_{(\left|X_{n 1}\right|>t)}\right]=0.
\end{equation*}
Then we know for any $\varepsilon>0$ and $\delta\in [0,1]$, there exists a $N$ such that for all $n \geq N$:
\begin{equation*}
\mathbb{P}\left(\left|
n^{-1} \sum_{k=1}^n X_{n k}-E X_{n1}
\right|>\varepsilon\right) \leq \delta .
\end{equation*}
\end{lemma}
\begin{proof}
We assumed uniform integrability for $\{P_n\}$, so for any fixed $\varepsilon$ and $\delta$, we can choose a $t$ such that 
\begin{equation*}
\sup _n E\left(\left|X_{n 1}\right| \mathbf{1}_{\left(\left|X_{n 1}\right|>t\right)}\right)<\varepsilon \delta / 6.
\end{equation*}
Let
\begin{equation}
\begin{aligned}
X_{nkt} & =X_{nk} \mathbf{1}_{\left(\left|X_{nk}\right| \leq t\right)} \\
Y_{nkt} & =X_{nk} \mathbf{1}_{\left(\left|X_{nk}\right|>t\right)}.
\end{aligned}
\end{equation}
By definition, we have $X_{nk}=X_{nkt}+Y_{nkt}$. Denote $S_n = \sum_{k=1}^n X_{n k}$, we also have
\begin{equation*}
\begin{aligned}
\frac{S_n}{n} & =\frac{1}{n} \sum_{k=1}^n X_{nkt}+\frac{1}{n} \sum_{k=1}^n Y_{nkt} \\
& =: U_{nt}+V_{nt}.
\end{aligned}
\end{equation*}
The target of interest can be bounded as
\begin{equation*}
\left|\frac{S_n}{n}-\mu_n\right| \leq\left|U_{nt}-\mu_{nt}\right|+\left|V_{nt}\right|+\left|\mu_{nt}-\mu_n\right|
\end{equation*}
where $\mu_n = E[X_{n1}] < \infty$ (implied by UI) and $\mu_{nt}=E[X_{n1t}]$.

\textbf{Step 1} We have
$$
E[|V_{nt}|]\leq \frac{1}{n} \sum_{k=1}^n E\left|Y_{n1t}\right|=E\left(|X_{n1}| \mathbf{1}_{(|X_{n1}|>t)}\right).
$$
By definition of $t$,
\begin{equation*}
\mathbb{P}\left(\left|V_{nt}\right|>\varepsilon / 3\right) \leq \frac{3 E\left|V_{nt}\right|}{\varepsilon} \leq \delta / 2.
\end{equation*}
\textbf{Step 2} We also have a bound on mean shifts
\begin{equation*}
\left|\mu_{nt}-\mu_n\right| = \left|E\left(Y_{n1t}\right)\right|<\varepsilon \delta / 6<\varepsilon / 3.
\end{equation*}
\textbf{Step 3} Using Chebychev’s inequality,
\begin{equation*}
\mathbb{P}\left(\left|U_{n t}-\mu_{n t}\right|>\varepsilon/3\right) \leq \frac{9}{n\varepsilon^2} \operatorname{Var}\left(X_{n1t}\right)\le \frac{9t^2}{n\varepsilon^2}.
\end{equation*}
We can combine the three steps: for any $\varepsilon, \delta$, we can find a $t = t(\varepsilon,\delta)$ such that for all $n \geq N=\left\lceil\frac{18 t^2}{\varepsilon^2 \delta}\right\rceil$,
\begin{equation*}
\mathbb{P}\left(\left|\frac{S_n}{n}-\mu_n\right|>\varepsilon\right) \leq \mathbb{P}\left(\left|V_{n t}\right|>\varepsilon / 3\right) + \mathbb{P}\left(\left|U_{n t}-\mu_{n t}\right|>\varepsilon / 3\right) \leq \delta.
\end{equation*}
\end{proof}

\begin{lemma}
\label{lemma: cross term}
Under the assumptions of \Cref{th: one step}. The ``cross-term'' 
\begin{equation*}
\begin{aligned}
  \mathcal{W}&:=\left(P_n^{(m)}-P^{(m)}\right)\left((X-Z)^{\top} v^{(-m)}+\phi^{(-m)}(X, Z)-(X-Z)^{\top} v-\phi(X, Z)\right)\\
& =o_P(n^{-1/2})\,.  
\end{aligned}
\end{equation*}
\end{lemma}
\begin{proof}
We first split $\mathcal{W}$ into two parts: an inner product term and a term involving the influence function:
    \begin{equation*}
    \begin{aligned}
     & \left(P_n^{(m)}-P^{(m)}\right)\left((X-Z)^{\top} v^{(-m)}+\phi^{(-m)}(X, Z)-(X-Z)^{\top} v-\phi(X, Z)\right) \\
     & = \left(P_n^{(m)}-P^{(m)}\right)\left\{(X-Z)^{\top}\left(v^{(-m)}-v\right)\right\} + \\
     & \quad \left(P_n^{(m)}-P^{(m)}\right)\left(\phi^{(-m)}(X, Z)-\phi(X, Z)\right).
    \end{aligned}
\end{equation*}
The first inner product term above is just:
\begin{equation*}
    \begin{aligned}
        &\left(P_n^{(m)}-P^{(m)}\right)\left\{(X-Z)^{\top}\left(v^{(-m)}-v\right)\right\}\\
        & = \left(\mu_X^{(m)}-\mu_X\right)^{\top}\left(v^{(-m)}-v\right) - \left(\mu_Z^{(m)}-\mu_Z\right)^{\top}\left(v^{(-m)}-v\right),
    \end{aligned}
\end{equation*}
which is $o_P(n^{-1/2})$ (\Cref{lemma: cross term is small}).


For the influence function terms, a similar argument also holds. We split the influence function into parts related to $X$ and $Z$ respectively and bound them separately. 

\begin{equation}
\label{eq: influence into two}
    \begin{aligned}
      & \left(P_n^{(m)}-P^{(m)}\right)\left(\phi^{(-m)}(X, Z)-\phi(X, Z)\right)\\
       & = \left(P_n^{(m)}-P^{(m)}\right)\left(w \phi_X^{(-m)}(X)-w \phi_X(X)\right)+\\
       &\left(P_n^{(m)}-P^{(m)}\right)\left((1-w) \phi_Z^{(-m)}(Z)-(1-w) \phi_Z(Z)\right).
    \end{aligned}
\end{equation}
Consider the parts involving $X$:
\begin{align*}
     \mathcal{W}_X& := \left(P_n^{(m)}-P^{(m)}\right)\left(w \phi_X^{(-m)}(X)-w \phi_X(X)\right)\\
& = (n_X+n_Z)^{-1}\sum_{X_i\in\mathcal{D}^{(m)}} \Delta \phi_X^{(-m)}(X_i) - E[\Delta \phi_X^{(-m)}(X_i) \mid \mathcal{D}^{(-m)}],
\end{align*}
where $\Delta \phi_X^{(-m)}(x) := \phi_X^{(-m)}(x) - \phi_X(x)$ is the difference between the estimated $\phi_X$ function and the truth.

Applying Chebyshev's inequality:
\begin{equation*}
    \begin{aligned}
        & \mathbb{P}\left(\left|\mathcal{W}_X\right| \geq \epsilon n^{-1 / 2}\right)\\
        & =E\left[\mathbb{P}\left(\left|\mathcal{W}_X\right| \geq \epsilon n^{-1 / 2} \mid \mathcal{D}^{(-m)}\right)\right]\\
        & \leq 2 \epsilon^{-2} n E\left[\operatorname{Var}\left(\mathcal{W}_X \mid \mathcal{D}^{(-m)}\right)\right]\\
        & \leq 2 \epsilon^{-2} n\left(n_X+n_Z\right)^{-2} n_X E\left[E\left[\left(\Delta \phi_X^{(-m)}\left(X\right)\right)^2 \mid \mathcal{D}^{(-m)}\right]\right]\\
        & \leq \epsilon^{-2} E\left[\left(\phi_X^{(-m)}(X)-\phi_X(X)\right)^2\right].
    \end{aligned}
\end{equation*}

Given the assumption that 
\begin{equation*}
\lim _{n \rightarrow \infty}
E\left[\left(\phi_X^{(-m)}(X)-\phi_X(X)\right)^2\right] = 0,
\end{equation*}
we know $\mathcal{W}_X$ is $o_P(n^{-1 / 2})$. A similar argument also holds for the term associated with $Z$ in \eqref{eq: influence into two}. This implies their summation $\mathcal{W}$ is also of order $o_P(n^{-1 / 2})$.
\end{proof}

\begin{lemma}\label{lemma: cross term is small}
    Let $\mu_X^{(m)}$ be the simple sample mean using $\mathcal{D}^{(m)}$ and $u^{(-m)}$ is a vector constructed from $\mathcal{D}^{(-m)}$. Suppose $\lambda_1(\Sigma_X) \leq C$ and $E\left\|u^{(-m)}-u_n\right\|^2 \rightarrow 0 \text { as } n \rightarrow \infty$ for some deterministic $u_n$. Then we have 
    \begin{equation*}
        \lim _{n \rightarrow \infty} \mathbb{P}\left(\left|\left(\mu_X^{(m)}-\mu_X\right)^{\top}\left(u^{(-m)}-u_n\right)\right| \geq \epsilon n^{-1 / 2}\right)=0 
    \end{equation*}
    for any $\epsilon >0$.
\end{lemma}
\begin{proof}
For any $\epsilon>0$, we have
\begin{equation*}
\begin{aligned}
& \mathbb{P}\left(\left|\left(\mu_X^{(m)}-\mu_X\right)^{\top}\left(u^{(-m)}-u\right)\right| \geq \epsilon n^{-1 / 2}\right) \\
& =E\left[\mathbb{P}\left(\left|\left(\mu_X^{(m)}-\mu_X\right)^{\top}\left(u^{(-m)}-u\right)\right| \geq \epsilon n^{-1 / 2} \mid \mathcal{D}^{(-m)}\right)\right] \\
& \leq 2 \epsilon^{-2} n \cdot E\left[\operatorname{Var}\left(\left(\mu_X^{(m)}-\mu_X\right)^{\top}\left(u^{(-m)}-u\right) \mid \mathcal{D}^{(-m)}\right)\right] .
\end{aligned}
\end{equation*}
By independence between the data folds, we have,
\begin{equation*}
\begin{aligned}
\operatorname{Var}\left(\left(\mu_X^{(m)}-\mu_X\right)^{\top}\left(u^{(-m)}-u\right) \mid \mathcal{D}^{(-m)}\right) & =\left(u^{(-m)}-u\right)^{\top}\left(\frac{1}{n_X} \Sigma_X\right)\left(u^{(-m)}-u\right) \\
& \leq \frac{1}{n_X} \lambda_1\left(\Sigma_X\right) \cdot\left\|u^{(-m)}-u\right\|^2 .
\end{aligned}
\end{equation*}
Therefore,
\begin{equation*}
\mathbb{P}\left(\left|\left(\mu_X^{(m)}-\mu_X\right)^{\top}\left(u^{(-m)}-u\right)\right| \geq \epsilon n^{-1 / 2}\right) \leq 2 \epsilon^{-2} \lambda_1\left(\Sigma_X\right) \cdot E\left\|u^{(-m)}-u\right\|^2 .
\end{equation*}
Finally, if $E\left\|u^{(-m)}-u\right\|^2 \rightarrow 0$, then the probability above converges to 0 .
\end{proof}

\begin{lemma}
\label{lemma: remainder term}
Under the assumptions of \Cref{th: one step}. The ``remainder-term'' 
\begin{equation*}
\mathcal{Z}:=P^{(m)}\left((X-Z)^{\top} v^{(-m)}+\phi^{(-m)}(X, Z)\right)
\end{equation*}
in \eqref{eq: basic decomposition for debiased} satisfies
\begin{equation*}
    \lim _{n \rightarrow \infty} \mathbb{P}\left(|\mathcal{Z}| \geq \epsilon n^{-1 / 2}\right)=0,
\end{equation*}
for any $\epsilon > 0$.
\end{lemma}
\begin{proof}

 We split the remainder into several terms that we will bound separately. Recall the notation $w = n_X/(n_X + n_Z)$:
\begin{equation}\label{eq: split remainder}
\begin{aligned}
&P^{(m)}\left((X - Z)^{\top} v^{(-m)} + \phi^{(-m)}(X, Z)\right)\\
&= (\mu_X - \mu_Z)^{\top} (v^{(-m)} - v) \\
&\quad + w\, s^{(-m)\top} P^{(m)} \left\{ (X - \mu_X^{(-m)})(X - \mu_X^{(-m)})^{\top} - \Sigma^{(-m)} \right\} v^{(-m)} \\
&\quad + (1 - w)\, s^{(-m)\top} P^{(m)} \left\{ (Z - \mu_Z^{(-m)})(Z - \mu_Z^{(-m)})^{\top} - \Sigma^{(-m)} \right\} v^{(-m)} \\
&= (\mu_X - \mu_Z)^{\top} (v^{(-m)} - v) + s^{(-m)\top} (\Sigma - \Sigma^{(-m)}) v^{(-m)} \\
&\quad + w\, s^{(-m)\top} (\mu_X - \mu_X^{(-m)})(\mu_X - \mu_X^{(-m)})^{\top} v^{(-m)} \\
&\quad + (1 - w)\, s^{(-m)\top} (\mu_Z - \mu_Z^{(-m)})(\mu_Z - \mu_Z^{(-m)})^{\top} v^{(-m)}.
\end{aligned}
\end{equation}

For the third term in the last line, we have:
\begin{equation}
    \begin{aligned}\label{eq: simple X term}
        & \left|s^{(-m)\top}\left(\mu_X-\mu_X^{(-m)}\right)\left(\mu_X-\mu_X^{(-m)}\right)^{\top} v^{(-m)}\right|\\
        & \leq \left|s^{(-m) \top}\left(\mu_X-\mu_X^{(-m)}\right)\right| \left|\left(\mu_X-\mu_X^{(-m)}\right)^{\top} v^{(-m)}\right|\\
        & \leq \left|s^{(-m) \top}\left(\mu_X-\mu_X^{(-m)}\right)\right| \left\|\mu_X^{(-m)}-\mu_X\right\| \\
        &\stackrel{(I)}{\lesssim} \left\|\mu_X-\mu_Z\right\|\left\|\left(\lambda_1^{(-m)} I_p-\Sigma^{(-m)}\right)^{+}\right\|\left\|\mu_X^{(-m)}-\mu_X\right\|^2\\
        & \stackrel{(II)}{\lesssim} \left\|\mu_X^{(-m)}-\mu_X\right\|^2 = o_P(n^{-1/2}).
    \end{aligned}
\end{equation}
In step $(I)$ we used the explicit form of $s^{(-m)}$ \eqref{eq: definition nuisance} and $\|\mu_X - \mu_X^{(-m)}\| \vee \|\mu_Z - \mu_Z^{(-m)}\| = o_P(1)$---therefore it is the population mean-difference that dominates. In step $(II)$, we used that $\left\|\left(\lambda_1^{(-m)} I_p-\Sigma^{(-m)}\right)^{+}\right\|$ is $O_P(1)$. This condition is verified in \Cref{lemma: bound terms in remainder}.

Similarly, the forth term in \eqref{eq: split remainder} can be bounded as:
\begin{equation}\label{eq: simple Z term}
\left|s^{(-m)\top}\left(\mu_Z-\mu_Z^{(-m)}\right)\left(\mu_Z-\mu_Z^{(-m)}\right)^{\top} v^{(-m)}\right| \lesssim o_P(n^{-1/2}).
\end{equation}

In the rest of the proof, we bound the first two terms in \eqref{eq: split remainder}, leveraging that the influence function corresponds to the first-order derivative of the target functional. Let $t \in[0,1]$. Define an interpolation matrix between the estimated covariance matrix and the population one:
\begin{equation*}
\Sigma_t= \Sigma^{(-m)}(1-t)+t \Sigma.
\end{equation*}
And define the eigenvector mapping $v:[0,1] \rightarrow \mathbb{R}^p$ as
\begin{equation*}
v(t)=\text { the first eigenvector of matrix } \Sigma_t \text {.}
\end{equation*}
We can see that $v(0)=v^{(-m)}$ and $v(1)=v$. Since the $v(t)$ and $-v(t)$ are eigenvectors of a matrix at the same time, we further require $v(t)^\top v(1) > 0$ for all $t$ to make this mapping well-defined. Similarly, we define the mapping $\lambda_1: [0,1]\rightarrow \mathbb{R}$ that returns the largest eigenvalue of matrix $\Sigma_t$.

Therefore,
\begin{equation}
\label{eq: debias taylor}
\begin{aligned}
v-v^{(-m)}
& =v(1)-v(0)  \stackrel{(I)}{=}\int_0^1 \frac{d v(t)}{d t} d t \\
& \stackrel{(II)}{=} \int_0^1 (\lambda_1(\Sigma_t)I_p - \Sigma_t)^+ \frac{d \Sigma_t}{d t} v(t) d t \\
& =\int_0^1 D_t\left(\Sigma-\Sigma^{(-m)}\right) v(t) d t \text {  denote } D_t=\left(\lambda_1\left(\Sigma_t\right) I_p-\Sigma_t\right)^{+}\\
& = D_0(\Sigma-\Sigma^{(-m)}) v(0)+\underbrace{\int_0^1(D_t-D_0)(\Sigma-\Sigma^{(-m)}) v(t) d t}_{B} \\
& +\underbrace{D_0(\Sigma-\Sigma^{(-m)}) \int_0^1(v(t)-v(0)) d t}_{C}
\end{aligned}
\end{equation}
In step $(I)$ and $(II)$ we use the derivative of $v$ exists and plug in its explicit form \cite{magnus1985differentiating, critchley1985influence}. Noting that $D_0 = (\lambda^{(-m)} I_p - \Sigma^{(-m)})^+$ and $v(0) = v^{(-m)}$, we multiply both sides of \eqref{eq: debias taylor} by $\left(\mu_X^{(-m)}-\mu_Z^{(-m)}\right)^{\top}$, we have:
\begin{align*}
    & \left(\mu_X^{(-m)}-\mu_Z^{(-m)}\right)^{\top}\left(v-v^{(-m)}\right) \\
    & = s^{(-m)\top}\left(\Sigma-\Sigma^{(-m)}\right) v^{(-m)} + \left(\mu_X^{(-m)}-\mu_Z^{(-m)}\right)^{\top}(B+C)\\
\Rightarrow \quad& s^{(-m)\top}\left(\Sigma-\Sigma^{(-m)}\right) v^{(-m)}\\
& = \left(\mu_X^{(-m)}-\mu_Z^{(-m)}\right)^{\top}(v - v^{(-m)}) - \left(\mu_X^{(-m)}-\mu_Z^{(-m)}\right)^{\top}(B + C).
\end{align*}
Go back to the first two terms in the last line of \eqref{eq: split remainder}:
\begin{equation*}
\begin{aligned}
& \left(\mu_X-\mu_Z\right)^{\top}\left(v^{(-m)}-v\right) + s^{(-m)\top}\left(\Sigma-\Sigma^{(-m)}\right) v^{(-m)}\\
& = \left(\mu_X^{(-m)}-\mu_X-\mu_Z^{(-m)}+\mu_Z\right)^{\top}\left(v-v^{(-m)}\right)-\left(\mu_X^{(-m)}-\mu_Z^{(-m)}\right)^{\top}(B+C)
\end{aligned}
\end{equation*}
Under our assumptions, the products above
\begin{equation}\label{eq: demon trio}
\begin{aligned}
&\left(\mu_X^{(-m)}-\mu_X-\mu_Z^{(-m)}+\mu_Z\right)^{\top}\left(v-v^{(-m)}\right)\\
   & \left(\mu_X^{(-m)}-\mu_Z^{(-m)}\right)^{\top}\int_0^1(D_t-D_0)\left(\Sigma-\Sigma^{(-m)}\right) v(t) d t\\
   & \left(\mu_X^{(-m)}-\mu_Z^{(-m)}\right)^{\top}D_0\left(\Sigma-\Sigma^{(-m)}\right) \int_0^1(v(t)-v(0)) d t
\end{aligned}
\end{equation}
 are all of order $o_P(n^{-1/2})$. We present the details of the argument in \Cref{lemma: bound terms in remainder}. Combine this result with \eqref{eq: simple X term} and \eqref{eq: simple Z term}, we conclude our proof.
\end{proof}

In the following lemma, we show the remainder terms are small under the conditions listed in the main text.
\begin{lemma}\label{lemma: bound terms in remainder}
Under the same assumptions as \Cref{th: one step}. The operator norms of the $D_t$ matrices---defined in \eqref{eq: debias taylor}---are all bounded by a constant with probability converging to $1$. Moreover, we know the three product terms in \eqref{eq: demon trio} are all of order $o_P(n^{-1/2})$.
\end{lemma}
\begin{proof}
We are going to bound the three terms one by one.
\paragraph{Part 1. Bound} $
\left(\mu_X^{(-m)}-\mu_Z^{(-m)}\right)^{\top} \int_0^1\left(D_t-D_0\right)\left(\Sigma-\Sigma^{(-m)}\right) v(t) d t
$.
\begin{equation}\label{eq: step1}
    \begin{aligned}
      & \left(\mu_X^{(-m)}-\mu_Z^{(-m)}\right)^{\top} \int_0^1\left(D_t-D_0\right)\left(\Sigma-\Sigma^{(-m)}\right) v(t) d t\\
    & = \int_0^1\left(\mu_X^{(-m)}-\mu_Z^{(-m)}\right)^{\top}\left(D_t-D_0\right)\left(\Sigma-\Sigma^{(-m)}\right) v(t) d t \\
    & \leq  \sup _{t \in[0,1]}\left\|\left(\mu_X^{(-m)}-\mu_Z^{(-m)}\right)^{\top}\left(D_t-D_0\right)\right\|\left\|\left(\Sigma-\Sigma^{(-m)}\right) v(t)\right\|.
    \end{aligned}
\end{equation}
Bounding the second term is straightforward:
\begin{equation}\label{eq: step2}
    \left\|\left(\Sigma-\Sigma^{(-m)}\right) v(t)\right\| \leq\left\|\Sigma-\Sigma^{(-m)}\right\| .
\end{equation}
Now we just need to handle the first one in \eqref{eq: step1}:
\begin{equation}\label{eq: diff shows up}
\sup _{t \in[0,1]}\left\|\left(\mu_X^{(-m)}-\mu_Z^{(-m)}\right)^{\top}\left(D_t-D_0\right)\right\| \leq 2\left\|\mu_X^{(-m)}-\mu_Z^{(-m)}\right\| \sup _{t \in[0,1]}\left\|D_t-D_1\right\|.
\end{equation}

We need the following perturbation result regarding the pseudo-inverse matrices from the literature:
\begin{theorem}(Theorem~3.3 in \cite{stewart1977perturbation}) For any matrices $\mathbb{A}$ and $\mathbb{B}$ with $\mathbb{B}=\mathbb{A}+\mathbb{F}$,
\begin{equation}
\left\|\mathbb{B}^{+}-\mathbb{A}^{+}\right\| \leq 
\frac{1+\sqrt{5}}{2}
\max \left\{\left\|\mathbb{A}^{+}\right\|^2,\left\|\mathbb{B}^{+}\right\|^2\right\}\|\mathbb{F}\|.
\end{equation}
\end{theorem}

Apply this theorem to our setting: for any $t\in[0,1]$,
\begin{equation}\label{eq: bound the pseudonorm difference}
    \begin{aligned}
            \left\|D_t-D_1\right\| & \lesssim 
\max \left\{\left\|D_t\right\|^2,\left\|D_1\right\|^2\right\}\|D_t^+ - D_1^+\|\\
& = \max \left\{\left\|D_t\right\|^2,\left\|D_1\right\|^2\right\}\|\lambda_1(\Sigma_t)I_p - \Sigma_t - \lambda_1(\Sigma_1)I_p + \Sigma_1\|\\
& \leq \max \left\{\left\|D_t\right\|^2,\left\|D_1\right\|^2\right\}\{|\lambda_1(\Sigma_t) - \lambda_1| + \|\Sigma_t - \Sigma_1\|\}\\
& \leq \max \left\{\left\|D_t\right\|^2,\left\|D_1\right\|^2\right\}\{|\lambda_1(\Sigma_t) - \lambda_1| + \|\Sigma^{(-m)} - \Sigma\|\}\\
& \stackrel{(I)}{\lesssim} \max \left\{\left\|D_t\right\|^2,\left\|D_1\right\|^2\right\}\|\Sigma^{(-m)} - \Sigma\|.
    \end{aligned}
\end{equation}

In step $(I)$, we applied Weyl's inequality to bound the difference between eigenvalues by the operator norm of the difference matrix. Specifically,
\begin{equation*}
\left|\lambda_1\left(\Sigma_t\right)-\lambda_1\right| \leq\left\|\Sigma_t-\Sigma\right\| \leq\left\|\Sigma^{(-m)}-\Sigma\right\| .
\end{equation*}
For a discussion and proof, see Section~8.1.2 of \cite{wainwright2019high}.

Now we are going to show the spectral norm of $D_t, D_1$ in \eqref{eq: bound the pseudonorm difference} are bounded with probability converging to 1 for any $t$. In fact (e.g., equation (3.3) in \cite{wedin1973perturbation}), the $\|\cdot\|$-norm of $D_t$ is equal to the inverse of the smallest (non-zero) singular value of $D_t^+ = \lambda_1(\Sigma_t)I_p - \Sigma_t$. A lower bound on the latter implies an upper bound on the operator norm of $D_t$. We proceed as follows: for any $j\in\{1,...,\text{rank}(D_t^+)\}$:
\begin{equation*}
    \begin{aligned}
      \sigma_{j}(D_t^+) 
      & = \sigma_{j}(\lambda_1I_p - \Sigma + D_t^+ - (\lambda_1I_p - \Sigma))\\
      & \geq \sigma_j(\lambda_1I_p - \Sigma) - \|D_t^+ - (\lambda_1I_p - \Sigma)\|\\
      & \geq \sigma_j(\lambda_1I_p - \Sigma) - 2\|\Sigma^{(-m)} - \Sigma\|\\
      & \geq (\lambda_1 - \lambda_2)- 2\|\Sigma^{(-m)} - \Sigma\|.
    \end{aligned}
\end{equation*}
So we know the smallest singular value can be lower bounded by $(\lambda_1 - \lambda_2)- 2\|\Sigma^{(-m)} - \Sigma\|$. Since we assumed the eigen-gap $\omega$ is greater than zero and $\|\Sigma^{(-m)} - \Sigma\|\rightarrow 0$ with probability converging to $1$, we conclude the $\|D_t\|^2$ term in \eqref{eq: bound the pseudonorm difference} can be bounded from above for large $n$. 

This implies
\begin{equation*}
    \sup_{t\in[0,1]}\|D_t - D_1\| \lesssim \|\Sigma^{(-m)} - \Sigma\|\text{ for large }n.
\end{equation*}
Combine it with \eqref{eq: step1}, \eqref{eq: step2} and \eqref{eq: diff shows up}:
\begin{equation}\label{eq: conclusion part 1}
    \begin{aligned}
        &\left(\mu_X^{(-m)}-\mu_Z^{(-m)}\right)^{\top} \int_0^1\left(D_t-D_0\right)\left(\Sigma-\Sigma^{(-m)}\right) v(t) d t\\
        & \lesssim \left\|\mu_X^{(-m)}-\mu_Z^{(-m)}\right\|\left\|\Sigma^{(-m)}-\Sigma\right\|^2 = o_P(n^{-1/2}).
    \end{aligned}
\end{equation}

\paragraph{Par 2. Bound} $\left(\mu_X^{(-m)}-\mu_Z^{(-m)}\right)^{\top} D_0\left(\Sigma-\Sigma^{(-m)}\right) \int_0^1(v(t)-v(0)) d t$.
\begin{equation}\label{eq: step 3}
    \begin{aligned}
        &\left(\mu_X^{(-m)}-\mu_Z^{(-m)}\right)^{\top} D_0\left(\Sigma-\Sigma^{(-m)}\right) \int_0^1(v(t)-v(0)) d t\\
        & \leq \left\|\left(\mu_X^{(-m)}-\mu_Z^{(-m)}\right)^{\top} D_0\right\|\cdot \sup _{t \in[0,1]}\left\|\left(\Sigma-\Sigma^{(-m)}\right)(v(t)-v(0))\right\|\\
        & \leq \left\|\left(\mu_X^{(-m)}-\mu_Z^{(-m)}\right)^{\top} D_0\right\| \left\|\Sigma-\Sigma^{(-m)}\right\| \sup_{t\in[0,1]}\|v(t)-v(0)\|.
    \end{aligned}
\end{equation}
We state the following version Davis-Kahan theorem to bound the difference between eigenvectors.

\begin{theorem}(A special case of Corollary~1 in \cite{yu2015useful})\label{th: eigenvector pertubation}
    Let $\mathbb\Sigma, \mathbb{\hat{\Sigma}} \in \mathbb{R}^{p \times p}$ be symmetric matrices. Assume the eigengap between the first two eigenvalues is strictly positive: $\mathbb{w} = \lambda_{1}(\mathbb\Sigma) - \lambda_{2}(\mathbb\Sigma) > 0$. If $\mathbb{v}, \mathbb{\hat{v}} \in \mathbb{R}^p$ satisfy $\mathbb\Sigma \mathbb{v}=\lambda_1(\mathbb\Sigma) \mathbb{v}$ and $\mathbb{\hat{\Sigma}} \mathbb{\hat{v}}=\lambda_1(\mathbb{\hat{\Sigma}})\mathbb{\hat{v}}$. Moreover, if $\mathbb{\hat{v}}^{\top} \mathbb{v} \geq 0$, then,
$$
\|\mathbb{\hat{v}}-\mathbb v\| \leq \mathbb{w}^{-1}2^{3 / 2}\|\mathbb{\hat{\Sigma}}-\mathbb{\Sigma}\|.
$$
\end{theorem}

In our case, the $\mathbb{\hat\Sigma}$ in \Cref{th: eigenvector pertubation} is $\Sigma_t=\Sigma^{(-m)}(1-t)+t \Sigma$. And we have the bound:
\begin{equation}\label{eq: step 4}
    \sup _{t \in[0,1]}\|v(t)-v(0)\| \lesssim \sup_{t \in [0,1]}\omega^{-1}\|(1-t)(\Sigma^{(-m)}-\Sigma)\| \leq \omega^{-1} \|\Sigma^{(-m)}-\Sigma\|.
\end{equation}
Combine \eqref{eq: step 3} and \eqref{eq: step 4}, then we know under our assumptions:
\begin{equation}\label{eq: conclusion part 2}
\begin{aligned}
  &\left(\mu_X^{(-m)}-\mu_Z^{(-m)}\right)^{\top} D_0\left(\Sigma-\Sigma^{(-m)}\right) \int_0^1(v(t)-v(0)) d t\\
  &\lesssim 
\left\|\left(\mu_X^{(-m)}-\mu_Z^{(-m)}\right)^{\top} D_0\right\|\omega^{-1}\left\|\Sigma-\Sigma^{(-m)}\right\|^2\\
& \lesssim \left\|\mu_X^{(-m)}-\mu_Z^{(-m)}\right\|\omega^{-1}\left\|\Sigma-\Sigma^{(-m)}\right\|^2\\
& = o_P(n^{-1/2}).  
\end{aligned}
\end{equation}

\paragraph{Part 3. Bound} $\left(\mu_X^{(-m)}-\mu_X-\mu_Z^{(-m)}+\mu_Z\right)^{\top}\left(v-v^{(-m)}\right)$.\\
This term is easy to handle given the results established above. The first half of the above quantity can be bounded as:
\begin{equation*}
\begin{aligned}
    &\left(\mu_X^{(-m)}-\mu_X\right)^{\top}\left(v-v^{(-m)}\right)\\
    & \leq \left\|\mu_X^{(-m)}-\mu_X\right\|\left\|v-v^{(-m)}\right\|\\
    & \stackrel{(I)}{\lesssim} \left\|\mu_X^{(-m)}-\mu_X\right\| \omega^{-1}\left\|\Sigma-\Sigma^{(-m)}\right\| = o_P(n^{-1/2}).
\end{aligned}
\end{equation*}
In step $(I)$, we used the bound on the eigenvectors \eqref{eq: step 4} with $t = 1$.
\end{proof}

\section{Proof of \Cref{th: approximate orthogonal}}
\label{app: app_orthogonal}
\begin{proof}[Proof of \Cref{th: approximate orthogonal}]
 For each $m\in[M]$, we have the following decomposition:
    \begin{equation}
\begin{aligned}
&\left(\mu_X^{(m)}-\mu_Z^{(m)}\right)^{\top} u^{(-m)}\\
& = \left(\mu_X^{(m)}-\mu_Z^{(m)}\right)^{\top} u^{(-m)} - \left(\mu_X-\mu_Z\right)^{\top} u^{(-m)} + \left(\mu_X-\mu_Z\right)^{\top} u^{(-m)}\\
& \stackrel{(I)}{ = }\left(\mu_X^{(m)}-\mu_Z^{(m)}\right)^{\top} u^{(-m)} - \left(\mu_X-\mu_Z\right)^{\top} u^{(-m)} + o_P(n^{-1/2})\\
& = \left(\mu_X^{(m)}- \mu_X\right)^\top u - \left(\mu_Z^{(m)} - \mu_Z\right)^{\top} u + \\
& \quad \left(\mu_X^{(m)}-\mu_X\right)^{\top}\left(u^{(-m)}-u\right)-\left(\mu_Z^{(m)}-\mu_Z\right)^{\top}\left(u^{(-m)}-u\right)+ o_P(n^{-1/2})\\
& \stackrel{(II)}{ = } \left(\mu_X^{(m)}-\mu_X\right)^{\top} u-\left(\mu_Z^{(m)}-\mu_Z\right)^{\top} u + o_P(n^{-1/2}).
\end{aligned}
\end{equation}
Step (I) uses approximate orthogonality \eqref{eq:3.2-approx-orth}. In step (II), we applied \Cref{lemma: cross term is small}.

Now, we examine the distribution of 
$\sum_{m \in[M]}\left(\mu_X^{(m)}-\mu_X\right)^{\top} u-\left(\mu_Z^{(m)}-\mu_Z\right)^{\top} u$. Define
\begin{equation}
\begin{aligned}
  Q & = \sigma_{\rm pi}^{-1}
\sum_{m \in[M]}\left(\mu_X^{(m)}-\mu_X\right)^{\top} u-\left(\mu_Z^{(m)}-\mu_Z\right)^{\top} u\\
& = \sigma_{\rm pi}^{-1}\left\{\sum_{i\in [N_X]} n_X^{-1}(X_i - \mu_X)^\top u + \sum_{i\in [N_Z]} n_Z^{-1}(Z_i - \mu_Z)^\top u\right\},
\end{aligned}
\end{equation}
with
\begin{equation}
    \sigma_{\rm pi} 
    = \sqrt{Mn_X^{-1} \operatorname{Var}\left(X^{\top} u\right)+Mn_Z^{-1}  \operatorname{Var}\left(Z^{\top} u\right)}.  
\end{equation}

We need to apply Lindeberg's CLT to establish the asymptotic normality of $Q$. In our case, we need: For all $\epsilon >0$:
\begin{equation}
\lim_{n\rightarrow\infty}M \sigma_{\mathrm{pi}}^{-2} n_X^{-1} E\left[\left\{\left(X_1-\mu_X\right)^{\top} u\right\}^2 \mathbb{1}_{\left\{\left|\left(X_1-\mu_X\right)^{\top} u\right|>\epsilon n_X\sigma_{\mathrm{pi}}\right\}}\right]= 0
\end{equation}
and
\begin{equation}
\lim_{n\rightarrow\infty}M \sigma_{\mathrm{pi}}^{-2} n_Z^{-1} E\left[\left\{\left(Z_1-\mu_X\right)^{\top} u\right\}^2 \mathbb{1}_{\left\{\left|\left(Z_1-\mu_Z\right)^{\top} u\right|>\epsilon n_Z\sigma_{\mathrm{pi}}\right\}}\right]= 0.
\end{equation}

Note that $E\left[\left\{\left(X_1-\mu_X\right)^{\top} u\right\}^2\right] = {\rm Var}(X^\top u)$ is finite, we can apply the Dominated Convergence Theorem to conclude 
\begin{equation}
    \lim_{n\rightarrow\infty} E\left[\left\{\left(X_1-\mu_X\right)^{\top} u\right\}^2 \mathbb{1}_{\left\{\left|\left(X_1-\mu_X\right)^{\top} u\right|>\epsilon n_X \sigma_{\mathrm{pi}}\right\}}\right]= 0.
\end{equation}
Also note that $\sigma_{\mathrm{pi}}^{-2} n_X^{-1}$ will not blow up. Therefore, we conclude $Q\rightarrow \mathcal{N}(0,1)$ in distribution.

Now we know 
\begin{equation}
  T_{\rm pi} = \sigma_{\rm pi} / \hat\sigma_{\rm pi} \cdot Q  +o_P(1).
\end{equation}
The cross-fitting variance estimator $\hat\sigma_{\rm pi}(u)$, defined in \eqref{eq: cross-fitting variance}, is a natural choice that does not require significant extra computation. We use $\sigma_{\mathrm{pi}} / \hat{\sigma}_{\mathrm{pi}} \rightarrow 1$ in probability and apply Slutsky's theorem to conclude that $T_{\rm pi}$ converges to a standard Gaussian.
    
\end{proof}

\section{Using Discriminant Vector as Projection Direction}

\begin{figure}[!tbp]
    \centering
    \includegraphics[width =0.8\linewidth]{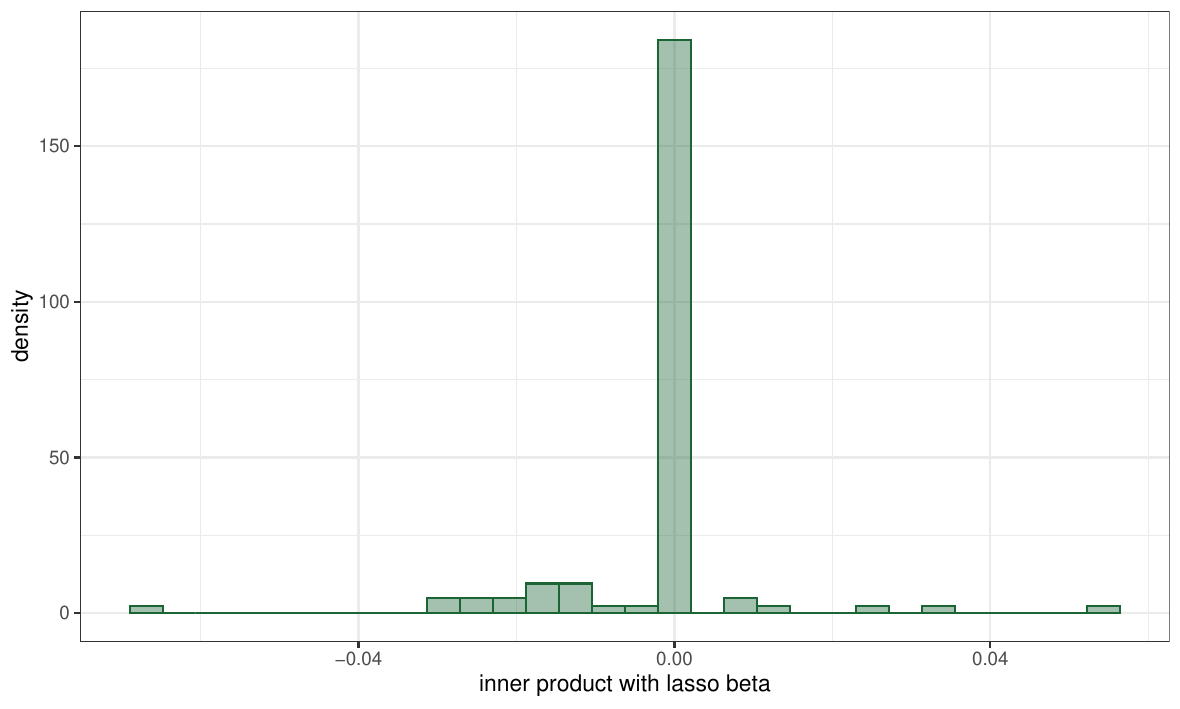}
    \caption{Degenerate distribution when directly projecting onto sparse estimates of discriminant direction. $N_X = 100, N_Z = 50, p =100$. Both $P_X,P_Z$ are normal distributions. Mean vectors are the same $\mu_X = \mu_Z$.}
    \label{appfig: funny lasso}
\end{figure}

In \Cref{section: anchored lasso} we mentioned the degeneracy when applying a sparse estimate of  discriminant direction (Lasso or LDA) directly as the projection direction. We present a simulated distribution of 
\begin{equation}
    \tilde T_{\rm deg} =M^{-1} \sum_{m=1}^M\left(\mu_X^{(m)}-\mu_Z^{(m)}\right)^{\top} \beta^{(-m)}
\end{equation}
in \Cref{appfig: funny lasso}, where the intermediate quantities are similarly calculated as in \eqref{eq:Lasso testing statistics}. Under the global null, cross-validated logistic Lasso vectors have a positive probability taking exactly zero (i.e. the tallest bar in the histogram is exactly zero rather than a very small number), indicating a non-Gaussian distribution of $\tilde T_{\rm deg}$.

\section{More Details on Simulated Data}

\subsection{Main Simulation Study}\label{app: simulated data}
We will use the notation that $a_s$ is a vector of length $s$ whose elements are all equal to $a\in\mathbb{R}$ and $I_p$ is an identity matrix of dimension $p\times p$.

We need to define a preliminary covariance matrix $\Sigma^{pre}$ to describe the ``normal part'' of the generating distribution. 
\begin{equation}
    \Sigma^{pre} = 100\cdot v_1v_1^\top + 50\cdot v_2v_2^\top + I_p
\end{equation}
where
\begin{equation}
\label{eq: zero inflated eigenvector}
    \begin{aligned}
        v_1 & = (1_{20}, 0_{980})^\top/\sqrt{20},\\
        v_2 & = (0_{20}, 1_{20}, 0_{960})^\top/\sqrt{20}.
    \end{aligned}
\end{equation}

We use the following scheme to generate the samples $X_i$ (group 2 samples $Z_j$ can be done similarly, replacing $\mu_X^{pre}$ by $\mu_Z^{pre}$):
\begin{enumerate}
    \item Draw a normally distributed sample $X_i^{pre}$ from $\mathcal N(\mu_X^{pre}, \Sigma^{pre})$. The mean vector $\mu_X^{pre}$ varies according to different settings---we will describe them later.
    \item Mask $X_i^{pre}$ with zeros: For each dimension of this preliminary sample, $X_{i,k}^{pre}, k = 1,...,p$, we generate an independent binary variable $X^{coin}\in\{0,1\}$ such that $pr(X^{coin} = 0) = pr(X^{coin} = 1) = 0.5$. If $X^{coin} = 0$, we change $X_{i,k}^{pre}$ to $0$. Otherwise, we do not modify $X_{i,k}^{pre}$. The resulting zero-inflated sample is our final observed $X_i$.
\end{enumerate}

It is possible to formally keep track of the first two moments of $X_i$ and $Z_j$. Specifically, denote $\Sigma = E[(X - \mu_X)(X - \mu_X)^\top] = E[(Z - \mu_Z)(Z - \mu_Z)^\top]$, we know:
\begin{equation}
     \Sigma_{ij} = \begin{cases}
\Sigma^{pre}/2 & \text{if } i = j \in\{1,..,p\} \\
\Sigma^{pre}/4 & \text{if } i \neq j \in\{1,..,p\}
\end{cases}
\end{equation}

The covariance matrix $\Sigma$ can be approximated by a rank-2 matrix. Denote the eigenvalues of it as $\lambda_1 \geq \lambda_2 \geq ...\lambda_{1000}$, we have:
\begin{equation}
\begin{aligned}
    &\lambda_1 = 26.75\\
    &\lambda_2 = 13.625\\
    &\lambda_3 = ...=\lambda_{21} = 1.75\\
    &\lambda_{22} = ... = \lambda_{40} = 1.125\\
    &\lambda_{41} = ... = \lambda_{1000} = 0.5.
\end{aligned}
\end{equation}

The first two eigenvectors of $\Sigma$ are still $v_1,v_2$ presented in \eqref{eq: zero inflated eigenvector}.

The means are more straightforward: $\mu_X = \mu^{pre}_X/2$, $\mu_Z = \mu^{pre}_Z/2$.

Now we present the details of each setting: global null, projected null, and alternative.

Under the global null $\mu_X = \mu_Z$, we set
\begin{equation}
        \mu_X^{pre}  = \mu_Z^{pre} = (1_{20}, 0_{980})^\top
\end{equation}

For the projected null case:
\begin{equation}
\begin{aligned}
    \mu_X^{pre}  & = (1_{20}, 0_{980})^\top \\
    \mu_Z^{pre}  & = (1_{20}, 5_{20}, 0_{960})^\top
\end{aligned}
\end{equation}
Under the above projected null setting, $(\mu_X - \mu_Z)^\top v_1 = 0$ whereas $(\mu_X - \mu_Z)^\top v_2 \neq 0$. 

Under the alternative, we chose:
\begin{equation}
\begin{aligned}
    \mu_X^{pre}  & = (1_{20}, 0_{980})^\top \\
    \mu_Z^{pre}  & = (1.2_{20}, 0.9_{20}, 0_{960})^\top
\end{aligned}
\end{equation}
To get more variety of the simulation, we purposely put more signal on the second eigenvector direction (mathematically, $|(\mu_X - \mu_Z)^\top v_1| < |(\mu_X - \mu_Z)^\top v_2|$). In this case, $v_1$ is not the optimal direction to project onto and we are curious about the behavior of the proposed estimators.

\subsection{Simulation in \Cref{fig: marginal and plug in}}\label{app: simulation truncated normal}

We first generate two multivariate normal distributions using the following means:
\begin{equation}
    \begin{aligned}
        \mu_X & = (1_{300})^\top\\
        \mu_Z & = (1_{10}, 2_{10}, 1_{280})^\top
    \end{aligned}
\end{equation}

Define $\Sigma_1=\left(\sigma_{i j}\right) \in \mathbb{R}^{10 \times 10}$ with $\sigma_{i i}=2$ and $\sigma_{i j}=1.8$ for $i \neq j$, and $\Sigma_2=$ $\left(\sigma_{i j}\right) \in \mathbb{R}^{10 \times 10}$ with $\sigma_{i i}=1$ and $\sigma_{i j}=0.6$ for $i \neq j$.
Let $\Sigma_X=\Sigma_Z \in \mathbb{R}^{300 \times 300}$ be a block diagonal matrix with 30 blocks. The first two blocks are $\Sigma_1$ and $\Sigma_2$, while the remaining blocks are identity matrices $I_{10}$.

After generating the data matrices, we randomly choose half of the entries to set to $0$. Moreover, we also shrink any value less than $0.5$ to $0$, including all negative values.

\section{More Details on Real Data Analysis}\label{app: immune cell results}

\subsection{Data Preprocessing in \Cref{section: clearys data}}\label{app: cleary_pre_processing}

For each given gene $j\in [p]$, we use $Y_{ij}$ to denote its expression level in cell $i$. The following procedure is done for each $j$ separately. We normalize $Y_{ij}$ using the formula
\begin{equation*}
    \tilde{Y}_{ij}:=Y_{ij} \exp \left(-\hat{\beta}_{jK} \mathbf{1}\{i \text { is in cell-cyle phase } K\}\right) / N_i,
\end{equation*}
where $N_i = \sum_{j=1}^p Y_{ij}$ is the library size. The coefficient $\hat \beta_{jk}$ is obtained using a Poisson regression using $Y_{ij}$ as the outcome, cell-cycle phase indicators as covariates and $N_i$ as the offset. In our case, the cells are in one of the three cell phases.

The package \texttt{CSCORE} takes gene expression counts as input and discovers correlated dimensions under a latent factor model. We use the original counts of the $2000$ control cells as input and obtain a covariance matrix for the latent variable. Using this matrix as input to \texttt{WGCNA}, we identified 19 gene modules. Approximately 1000 genes, showing weak empirical correlations, were not assigned to any module. Based on biological function, we further divided the remaining genes into 23 modules, resulting in a total of 42 modules. Module sizes range from 12 to 129 genes, with an average size of 47.

\subsection{Gene Ontology in \Cref{fig: cleary_loading_concentration}}\label{app: GOID}

In the main text, we presented the GO interpretations for each gene module using descriptive terms for ease of reading. For reference, we provide the corresponding GO identifiers (GO IDs) here so the readers can locate the precise entries in public databases.

\begin{table}[!t]
\centering
\begin{tabular}{ll}
\hline
\textbf{Description} & \textbf{GO ID} \\
\hline
Cell-cell signaling & GO:0007267 \\
Leukocyte migration & GO:0050900 \\
Signaling receptor binding & GO:0005102 \\
Growth factor activity & GO:0008083 \\
Response to virus & GO:0009615 \\
Response to biotic stimulus & GO:0002831 \\
Innate immune response & GO:0045088 \\
Cytokine-mediated pathway & GO:0019221 \\
Response to biotic stimulus & GO:0002833 \\
Chemokine response & GO:1990868 \\
GPCR signaling & GO:0007186 \\
Cell homeostasis & GO:0019725 \\
\hline
\end{tabular}
\caption{GO ID for presented pathways.}
\label{tab:short_desc_go_ids}
\end{table}

\subsection{Supplement Results for the Lupus Study}

In the main text \Cref{section: real data}, we presented the support gene results for T4 cells. In this section, we also provide the analysis results for the other three types of immune cells in \Cref{appfig: upset t8} - \ref{appfig: upset cm}.

\begin{figure}[!htbp]
    \centering
    \includegraphics[width =0.9\linewidth]{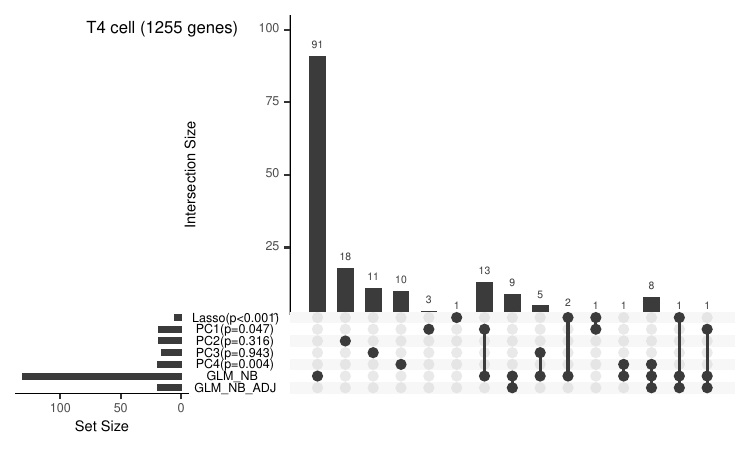}
    \caption{Upset plot, T4 cells.}
    \label{appfig: upset t4}
\end{figure}

\begin{figure}[!htbp]
    \centering
    \includegraphics[width =0.9\linewidth]{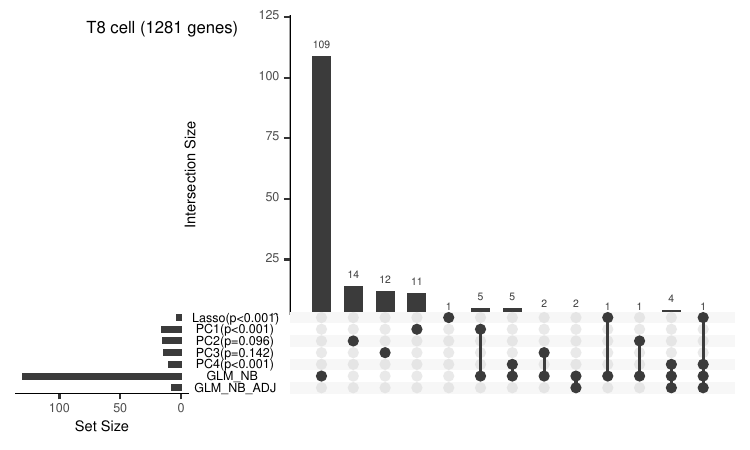}
    \caption{Upset plot, T8 cells.}
    \label{appfig: upset t8}
\end{figure}

\begin{figure}[!htbp]
    \centering
    \includegraphics[width =0.9\linewidth]{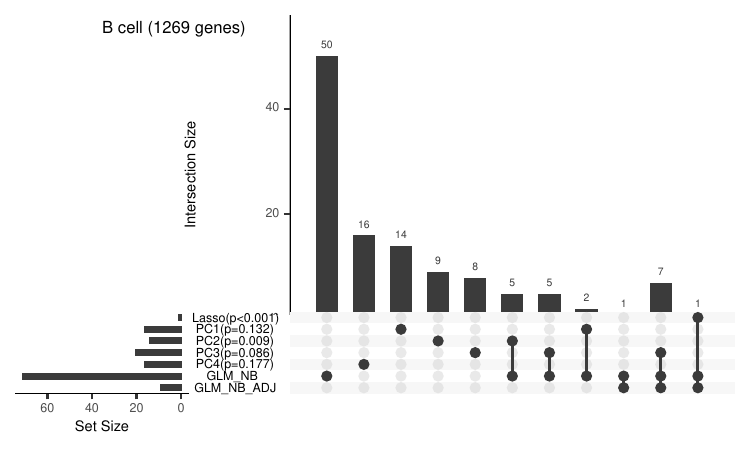}
    \caption{Upset plot, B cells.}
    \label{appfig: upset b}
\end{figure}

\begin{figure}[!htbp]
    \centering
    \includegraphics[width =0.9\linewidth]{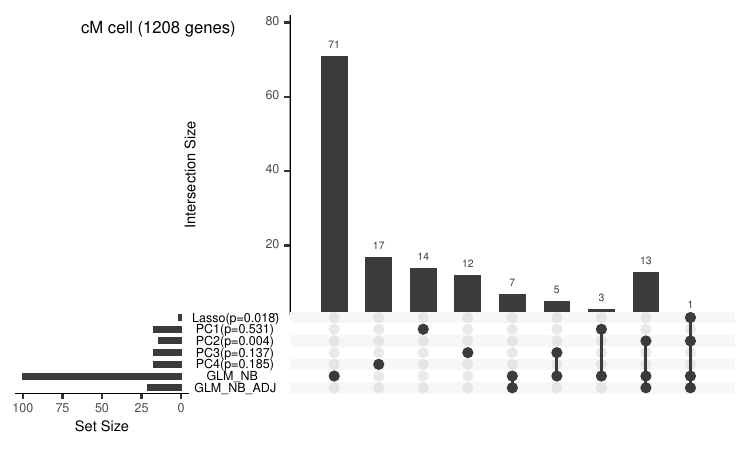}
    \caption{Upset plot, cM cells.}
    \label{appfig: upset cm}
\end{figure}

\begin{figure}[!htbp]
    \centering
    \includegraphics[width =\linewidth]{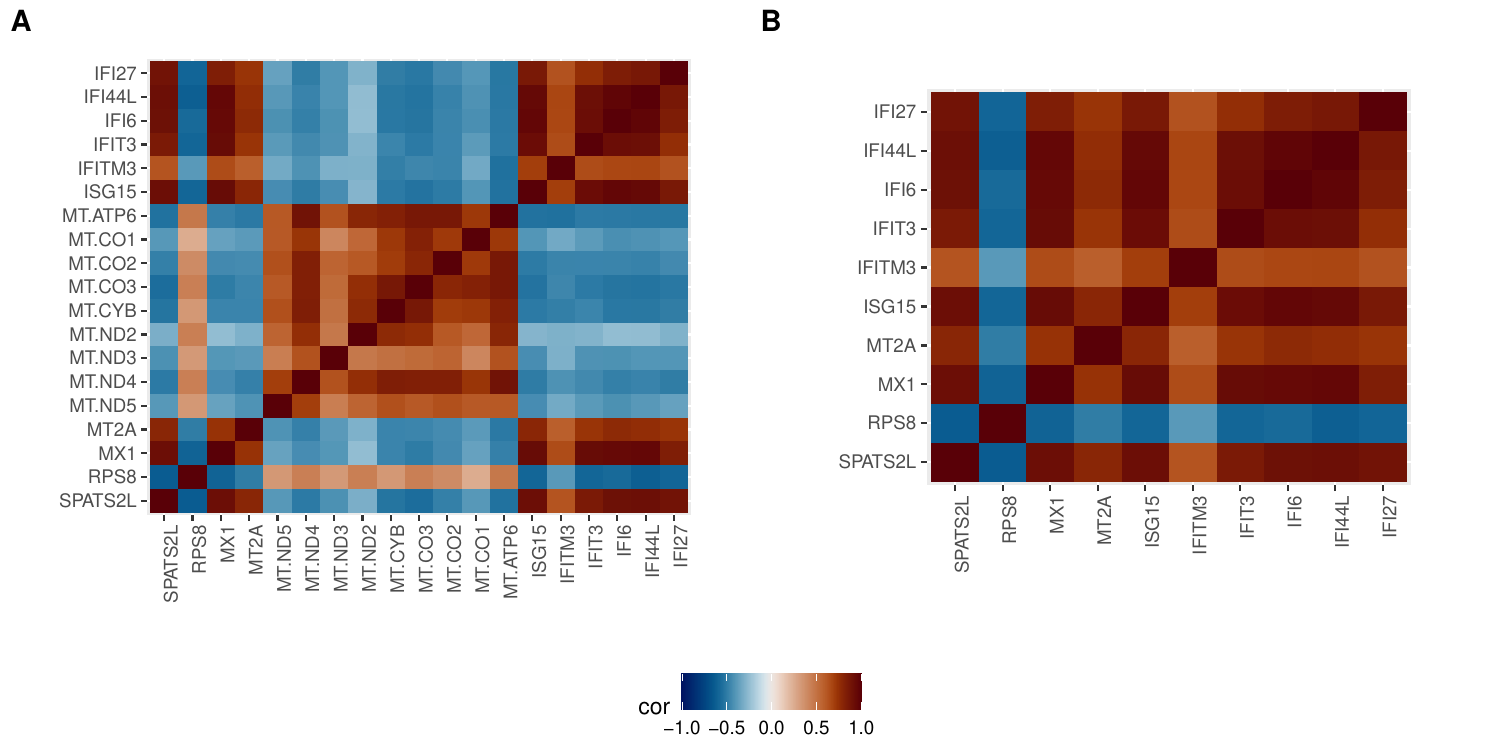}
    \caption{Heatmap plot for PC4, T4 cell. \textbf{A}. All 19 active genes. \textbf{B}. Removing the $9$ mitochondrial genes.}
    \label{appfig: inspect PC4}
\end{figure}

\newpage 
In \Cref{appfig: inspect PC4}, we give a zoomed-in assessment of PC4 support genes (panel A). If one were only interested in protein-encoding genes, the mitochondria genes would have been removed from the analysis, which would give a visually different correlation block (panel B). 

\end{document}